\documentclass[12pt]{article}
\usepackage{graphicx}
\usepackage{amssymb}
\usepackage{cite}    
\usepackage{bm}

\def\fun#1#2{\lower3.6pt\vbox{\baselineskip0pt\lineskip.9pt
\ialign{$\mathsurround=0pt#1\hfil##\hfil$\crcr#2\crcr\sim\crcr}}}

\renewcommand{\vec}[1]{{\bf #1}}
\newcommand{\vecg}[1]{\mbox{\boldmath $#1$}}

\newcommand{\be}{\begin{eqnarray}}
\newcommand{\ee}{\end{eqnarray}}
\newcommand{\bd}{\begin{displaymath}}
\newcommand{\ed}{\end{displaymath}}
\newcommand{\ba}{\begin{array}}
\newcommand{\ea}{\end{array}}
\newcommand{\bt}{\begin{tabular}}
\newcommand{\et}{\end{tabular}}

\newcommand{\Tr}{{\rm Tr\,}}
\newcommand{\dr}{{\rm d}}
\newcommand{\re}[1]{Eq.\,(\ref{#1})}

\begin{document}

\begin{titlepage}

\begin{flushright}

FTPI/MINN-04/21\\
UMN-TH-2310/04\\
hep-th/0405142

\end{flushright}

\vspace{0.5cm}

\begin{center}
{\large\bf Background field calculations and nonrenormalization theorems\\[1mm]
 in 4d  supersymmetric gauge theories\\[2mm]
 and their low--dimensional descendants}

\vspace{1.0cm}

{\bf Andrei Smilga}\\[1mm]
{\it SUBATECH, Universit\'e de
Nantes,\\  4 rue Alfred Kastler, BP 20722, Nantes  44307, France}\,\footnote{\,On 
leave of absence from ITEP, Moscow, Russia }\\[1mm]
and \\[1mm]
 {\bf Arkady Vainshtein} \\[1mm]
{\it William I.  Fine Theoretical Physics Institute, University of Minnesota,\\ 
116 Church St. S.E., Minneapolis, MN 55455, USA} \\

\end{center}

\vspace{1.5cm}

\begin{abstract}
 We analyze the structure of multiloop supergraphs contributing to the effective
Lagrangians in 4d supersymmetric gauge theories and in the models obtained from them by
dimensional reduction. When $d=4$, this gives the renormalization of the effective charge.
For $d < 4$, the 
low-energy effective Lagrangian describes the metric on
 the moduli space of classical vacua. These two problems turn out
to be closely related. In particular, we establish the relationship between the 4d 
nonrenormalization theorems (in minimal and extended supersymmetric theories) and their
low--dimensional counterparts.

\end{abstract}
 
\end{titlepage}

\newpage
\tableofcontents
\newpage

\section{Introduction}
Soon after the discovery of supersymmetry, it was understood that it imposes stringent
constraints on renormalization pattern of 4d field theories. 
In particular, there is no renormalization of a superpotential  ${\cal W}$ which is a 
function of the chiral fields $\Phi_{i}(x,\theta)$.  It is the so called
 $F$ term in the Lagrangian given by the the integral over the chiral subspace of superspace, 
$\int {\rm d}^{2}\theta  \,{\cal W}$.)

The gauge coupling term is also given by the integral over the chiral subspace 
of gauge invariant quantity, $\int {\rm d}^{2}\theta  \,\Tr {W}^\alpha W_{\alpha}$, 
what suggests its nonrenormalization.
It is well known that the situation is more complicated in this case.
 There is no, indeed, charge renormalization in ${\cal N}\! = \! 4$ super--Yang--Mills (SYM)
theory. In ${\cal N}\!  = \! 2$ theories, only 1--loop contribution in the $\beta$ function
survives.  In ${\cal N}\!  = \! 1$ theories,  multiloop contributions to the $\beta$ function
are related to renormalization of $Z$--factors, which allows one to evaluate higher loops
exactly in pure SYM theory and express them via anomalous dimensions of the matter
fields in the theories involving chiral matter multiplets \cite{NSVZ}.

The simplest way to prove all the listed nonrenormalization theorems is to analyze
the structure of the relevant supergraphs. We refer the reader to the textbooks 
\cite{WB,Siegel,West} for the proof of nonrenormalization theorems for superpotential, but recall
in some more details (following Refs.\cite{QED}) how it is done for gauge couplings, 
the subject of our interest here.

\begin{figure}[h]
   \begin{center}
 \includegraphics[width=4.0in]{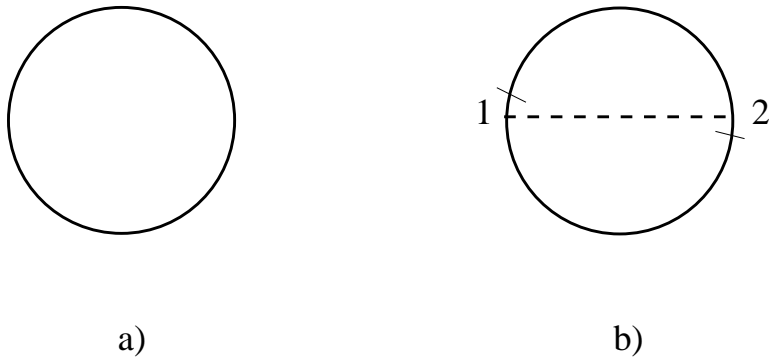}
        \vspace{-2mm}
    \end{center}
\caption{\small Contributions to the effective action, a) one loop; b) two loops. Solid lines 
are chiral field 
superpropagators $\langle \Phi_1 \bar \Phi_2 \rangle $ and the dashed line stands 
for vector superfields.  
The bar on the solid line marks
the $\bar \Phi$ end.}
\label{sgraph}
\end{figure}

Consider for simplicity Abelian theory --- supersymmetric electrodynamics. 
It involves the massless photon and photino described  the vector superfield $V$ 
and massive charged particles of spin 0 and 1/2 described by two chiral superfields 
$\Phi^{i}=\{S,T\}$ with opposite electric charges. To relate the physical 
charge $e_{\rm phys}$ measured in infrared, i.e. below the matter masses, to the bare 
charge $e_0$ defined at ultraviolet scale $\Lambda_{\rm UV}$, we have to 
evaluate the  supergraphs describing vacuum loops in the presence of a soft background gauge
field $V$. The relevant 1--loop and 2--loop graphs are depicted in Fig.\,\ref{sgraph}.\\[1mm]

The one--loop graph gives a nonvanishing correction to effective Lagrangian of the vector 
field $V$ at momenta much smaller than the mass of the matter fields,
 \be
\label{1loopSQED}
{\cal L}_{\rm eff}^{\rm 1-loop}(V) = \ {\rm Re}
\left\{ \left[\frac{1}{e^{2}_{0}}+\frac{1}{4\pi^{2}} 
\ln \frac{\Lambda_{\rm UV}} {m_0}\right]\frac{1}{2} \int\! {\rm d}^2\theta\, W^2 \right\}\,,
 \ee
where $e_{0}$ and $m_0$ are the bare charge and mass and the ultraviolet cut off 
$\Lambda_{\rm UV}$ is introduced as a mass of Pauli-Villars regulators.

The statement is that the two--loop graph and also higher loop contributions to ${\cal L}_{\rm eff}$
vanish identically. Indeed, according to the supergraph Feynman rules \cite{Siegel,West}, 
each vertex
involves the integral $\int {\rm d}^8z = \int {\rm d}^4x  \, {\rm d}^4\theta $ 
and the whole contribution
of the graph in Fig.\,\ref{sgraph}b is
$\Delta {\cal L}_{\rm eff}(V) = \frac 1{e^2} \int \dr^4\theta_1  {\cal K}^{(2)}(z_{1})$, where
$$ 
{\cal K}^{(2)}(z_{1}) = \frac {ie^2} 2\sum_{i,j=1,2}\int {\rm d}^8z_2 \langle \Phi_1^{i} 
\bar \Phi_2^{j} \rangle  \langle  \Phi_2^{j} \bar \Phi_1^{i}  \rangle
\langle v_1 v_2 \rangle \ .
$$   
 Here $\Phi^{i}$ stands for the charged chiral superfield $S$ or $T$ and the  
vector superfield $v$ 
is a quantum deviation from the classical background $V$. 
Furthermore, $\langle \Phi_1^{i} \bar \Phi_2^{j} \rangle, \ 
\langle v_1 v_2 \rangle$  are quantum superpropagators evaluated in external background 
$V$ and subscripts 1,2 refer to the superspace coordinates $z_{1,2}$.  Now, 
$ \langle v_1 v_2 \rangle$ does not depend
on external field and on its gauge. The charged field propagators are gauge--dependent:
\be
\langle \Phi_1^{i} \bar \Phi_2^{j} \rangle \to e^{-iq_{i}\Lambda_1} \langle \Phi_1^{i} 
\bar \Phi_2^{j} \rangle e^{iq_{j}\bar\Lambda_2}\,,
\ee
where $q_{i}=\{1,-1\}$ are electric charges of the fields $\Phi^{i}$.

The point is, however, that the {\it integrand} ${\cal K}^{(2)}$ is gauge--independent 
and should thereby 
be {\it locally}\,\footnote{\,Locality follows from the presence of an infrared cutoff 
(nonzero mass) in the theory. 
We postpone the discussion of what happens in massless theories till Sect.\,2.5.}
expressed via the gauge--invariant superfield $W_\alpha$. 
But $W_\alpha$ is a chiral superfield and
the integral over ${\rm d}^4\theta$ of any function of $W_\alpha$ vanishes. Therefore 
$\int {\rm d}^4\theta  \,{\cal K}^{(2)} =0$
Q.E.D. The same reasoning apply also to an arbitrary multiloop graph. 

Another way to prove the same statement is based on the fact that the effective
charge is a holomorphic function of $m_0$ \cite{SV}. Indeed, the higher powers of $\log m$
produce dependence of the effective charge on the phase of $m_{0}$, i.e., on the vacuum $\theta$ angle  -- the effect which does not occur in perturbation theory where terms presenting total derivatives do not contribute.

We hasten to comment that this does {\it not} mean that multiloop contributions to $\beta$ 
function in ${\cal N} =1$ supersymmetric QED vanish. Higher loops appear when expressing
the bare mass $m_0$ entering Eq.\,(\ref{1loopSQED}) via the physical mass $m_{\rm phys}$. 
The physical
mass {\it is} renormalized in spite of the fact that the mass term in the Lagrangian is not. 
Indeed,
the physical mass can be defined as the pole of the fermion propagator 
$\propto 1/(Z\!\!\!\not\!p - m_0)$, where $Z$
describes the renormalization of the {\it kinetic} term 
\be
Z \int {\rm d}^4\theta \left( \bar S e^V S + \bar T e^{-V} T \right) \ . 
\label{Zfac}
\ee
We have $m_0 = Z  m_{\rm phys}$ what leads to an exact relation expressing the charge 
renormalization via  the matter  $Z$ factor,
   \be
\label{ephys1}
\frac 1{e_{\rm phys}^2} \ =\  \frac 1{e_0^2} +\frac 1{4\pi^2} \ln \frac 
{\Lambda_{\rm UV}}{m_{\rm phys}} 
-\frac 1{4\pi^2} \ln Z\,.
\ee
In particular, using knowledge of $Z$ at one--loop level,
 \be
  \label{mrenorm}
Z=\frac{m_0}{m_{\rm phys}} = 1 -  \frac {e_0^2}{4\pi^2}  \ln \frac{\Lambda_{\rm UV}} 
{m_{\rm phys}} + \ldots \,,
  \ee
we obtain the two--loop renormalization of the charge
   \be
\label{ephys}
\frac 1{e_{\rm phys}^2} &=&  \frac 1{e_0^2} + 
\frac 1{4\pi^2} \ln \frac {\Lambda_{\rm UV}}{m_{\rm phys}} - 
\frac 1{4\pi^2} \ln \left[1 - \frac {e_0^2}{4\pi^2} \ln \frac {\Lambda_{\rm UV}}{m_{\rm phys}}  
\right]
\nonumber \\[1mm]
&=&\  \frac 1{e_0^2} + 
\frac 1{4\pi^2} \ln \frac {\Lambda_{\rm UV}}{m_{\rm phys}} 
+ \frac {e_0^2}{16\pi^4} \ln \frac {\Lambda_{\rm UV}}{m_{\rm phys}} + \ldots \ .
   \ee
    
In case of     
${\cal N}\! =\! 2$ supersymmetric electrodynamics, the above consideration shows an absence of 
higher loops because $Z=1$ in this case. Indeed, the ${\cal N}\! =\! 2$ SQED involves an extra 
neutral chiral superfield $\Upsilon$. Besides the $\Upsilon$ kinetic term, the Lagrangian 
contains the superpotential term $\propto \int {\rm d}^2\theta \, \Upsilon ST $.  The latter 
is not renormalized: this is the standard
$F$ term nonrenormalization theorem. The point is that this superpotential term is related
by extended supersymmetry to the charged field kinetic term. Hence, nonrenormalization of the
superpotential {\it implies} in  ${\cal N} = 2$ theory nonrenormalization of the kinetic term. 
In other words, in ${\cal N}\! =\! 2$ theory,
$m_{\rm phys} = m_0$ and hence only the 1-loop term in the $\beta$ function survives. 
It is valid for non-Abelian case as well. Moreover, if the matter content of the particular  
${\cal N}\! =\! 2$ theory is such that the one-loop $\beta$ function
vanishes, the theory is finite. The ${\cal N} = 4$ SYM theory belongs to this class.

The ${\cal N}\! =\! 2$ SQED represents a new phenomenon --- an infinite degeneracy of 
vacuum states. 
Indeed, different vacua of the theory are characterized by a value of $\Upsilon$ which serves 
as a modulus parameter in the set (moduli space) of vacua. In nonsupersymmetric 
theories, the vacuum degeneracy is always associated with spontaneous breaking of a continuous 
global symmetry what implies physical equivalence of different vacua --- chiral description 
 for pions 
in QCD is an example.
It is not the case for moduli spaces in supersymmetric theories. 
In  the ${\cal N}\! =\! 2$ SQED different values of $\Upsilon$ imply different masses of the 
charged field, i.e. different physics. The effective low-energy Lagrangian of this theory has 
the following form:
  \be
  \label{LeffUps}
{\cal L}_{\rm eff}=\frac{1}{2}\, {\rm Re}\left\{\int\! {\rm d}^2\theta
\left[\frac{1}{e^{2}_{0}}+\frac{1}{4\pi^{2}} \ln \frac{\Lambda_{\rm UV}} {\Upsilon}\right]W^2 
+\int\! {\rm d}^4\theta
\left[\frac{1}{e^{2}_{0}}+\frac{1}{4\pi^{2}} \left(\ln \frac {\Lambda_{\rm UV}} 
{\Upsilon}+1\right)\right]\bar\Upsilon \Upsilon\right\},
\ee
where we normalized $\Upsilon$ in such a way that its background value is equal to the mass 
of the charged field.\footnote{\,The bare mass $m_{0}$ provides just a shift in $\Upsilon$.}
For the lowest component $\upsilon$ of the modulus field $\Upsilon$ it gives
\be 
{\cal L}_{\rm eff}=
\frac{1}{2e^{2}(\upsilon)}\,
\partial_{\mu}\bar \upsilon \,\partial^{\mu}\upsilon \ =\ \frac 12 
\left[\frac{1}{e^{2}_{0}}+\frac{1}{8\pi^{2}}\,\ln \frac {|\Lambda_{\rm UV}|^{2}} 
{|\upsilon|^{2}}\right]
\partial_{\mu}\bar \upsilon \partial^{\mu}\upsilon\,,
\ee
what can be viewed as a metric in moduli space. We see a remarkable relation between the moduli  
metric and the effective charge renormalization in perturbation theory.

This relationship allows one to construct an alternative proof of the nonrenormalization theorem.
Actually, extended supersymmetry {\it dictates} the coefficient of $W^2$ in the first term in
(\ref{LeffUps}) to be a holomorphic function of $\Upsilon$. This generalizes the
comment above referring to holomorphic dependence on $m_0$ in ${\cal N} =1$ theories. Nonvanishing 
higher--order corrections would spoil this holomorphy and are not allowed. We
will discuss this in more details in Sect.\,3. 

For non-Abelian theory such as ${\cal N}\! =\! 2$ pure SYM theory with the SU(2) gauge group 
the low energy effective Lagrangian involves only Abelian degrees of freedom and 
has a similar form. In perturbative calculations, the massive matter fields in the
loops are  substituted 
in this case by the charged massive vector fields  $W^{\pm}_{\mu}$ 
and their superpartners. The coefficient of the logarithm is of the opposite 
sign, of course,  reflecting  asymptotic freedom at short distances,
   \be
  \frac{1}{g^{2}(\upsilon)}=
  \frac{1}{g^{2}_{0}}-\frac{1}{4\pi^{2}}\,\ln \frac {|\Lambda_{\rm UV}|^{2}} {|\upsilon|^{2}}=
  \frac{1}{4\pi^{2}}\,\ln \frac {|\upsilon|^{2}}{|\Lambda_{\rm IR}|^{2}}\,.
   \ee
Moreover, in this case all nonperturbative terms which are  powers of 
$|\Lambda_{\rm IR}|^{4}/|\Upsilon|^{4}$ are also known thanks to the Seiberg-Witten exact 
solution\cite{SeiWit}.  
Nonperturbative effects in moduli dynamics are even more crucial in case of 
${\cal N}\! =\! 1$ gauge theories where they could lead to appearance of superpotential for 
moduli\cite{instsuppot}.

Similar questions can be posed and solved for low-dimensional descendants, i.e.  theories 
obtained by dimensional reduction of the corresponding 4d theories (see \cite{sestry} for a recent
review). 
Typically, the Coulomb branch moduli space is enhanced in the descendants
compared to the 4d case, involving now the components of 4d vector potential in the 
reduced spatial directions.\footnote{\,The term {\it Coulomb branch} was coined for $d=4$ 
where it involves besides scalar fields also
massless photon mediating Coulomb interaction. Being reluctant  to invent new words, 
we will use this term for flat directions associated with the gauge
field and/or its superpartners also for $d=2$ and $d=1$ where the only reminiscent  of 
Coulomb phase is the absence of mass.}
E.g., in 4d ${\cal N}\!=\!1$ theories, there is no Coulomb branch whatsoever, but it 
appears in lower dimensions. In particular, reduction of the ${\cal N}\!=\!1$ SYM to one 
dimension leads to the effective Lagrangian representing  a nonstandard ``symplectic'' 
${\cal N}\! =\! 2$ $\sigma$ model defined
on a $3r$--dimensional target space ($r$ is the rank of the group) 
with certain conditions for the metric \cite{LeffN1,eshche}.
The 2d effective Lagrangian is a K\"ahler $\sigma$ model living on 
target space of  $r$ complex dimensions.

If we start from ${\cal N}\! =\!2$ SYM in  4 dimensions, the effective Lagrangians become 
fancier and prettier.
The 1d effective Lagrangian represents then
\cite{N2} a generalization of the
$\sigma$ model suggested in \cite{DE} with $(3+2)r = 5r$ dimensional target 
space.\footnote{\,The fermions belong to the spinor representation of SO(5)$\equiv Sp(2)$
and one can call it the symplectic model of the second kind. For the symplectic models of the first
kind with 2 complex supercharges, the fermions are doublets in SO(3)$\equiv Sp(1)$.}
The 2d effective Lagrangian \cite{DS,N2}
belongs to the class of twisted ${\cal N}\! =\!4$ $\sigma$ models \cite{GHR}. Finally, 
3d effective Lagrangians are
hyper--K\"ahler $\sigma$ models. For each unit of $r$, the moduli space
involves 2 variables coming from adjoint scalar, one variable from the component
of the vector potential in the reduced dimension and one variable 
representing a dual 3d photon.
The moduli spaces represent  Atiyah--Hitchin (AH) manifolds for 
unitary groups \cite{SW+CH} and
 hyper--K\"ahler manifolds obtained from AH manifolds after certain factorizations 
for other simple Lie groups \cite{seliv}.

The metric on the target spaces of these $\sigma$ models can be 
determined by evaluating perturbative loop corrections to the effective Lagrangian. 
The problem is conceptually very similar to that of effective
charge renormalization in $d\!=\!4$. Indeed,  as was noted in \cite{Akhmedov}, the 1--loop 
corrections to the metric in lower dimensions are rigidly related
to the 1--loop coefficient in the 4--dimensional $\beta$ function.
 
Nonrenormalization theorems can also be formulated in lower dimensions. In parti\-cular, 
for the descendants of
 ${\cal N}\! = \!2$ theories, all higher loop corrections to the metric beyond one loop vanish.
This is very similar to what happens in 4 dimensions. The  proof of the 
low--dimensional nonrenormalization theorems is  based  on the constraints
on the form of the effective Lagrangian following from supersymmetry. For $d=4$, such constraints
lead to holomorphic dependence on moduli. In lower dimensions, a generalization
of this is harmonic dependence on extended moduli. For example, 
for the groups of rank 1, the presence of 4 complex
supercharges in the twisted $\sigma$ model ($d\!=\!2$) as well as in the Diaconescu--Entin model 
($d\!=\!1$) requires
that the metric represents a harmonic O(4) [O(5)] invariant function. With 
\be
{\cal L}_{\rm eff}=\frac{1}{2e^{2}}\,h(\vec A)\, \partial_{\mu}{\vec A} \partial^{\mu}{\vec A}
\ee
the only possibility is
\be
\label{hN2}
&& d=2: \qquad {\vec A}\in{\mathbb R}^{4}\,,\qquad h(\vec{A}) = 1 +  \frac C{\vec{A}^2}\,;
\nonumber \\[1mm]
&& d=1: \qquad {\vec A}\in{\mathbb R}^{5}\,,\qquad h(\vec{A}) = 1 +  \frac C{|\vec{A}|^3}\,.
  \ee
But this means that the would be higher loop corrections to the metric 
$\Delta h^{l-{\rm loops}}_{d=2} \propto 1/|\vec{A}|^{2l}$  and 
$\Delta h^{l-{\rm loops}}_{d=1} \propto 1/|\vec{A}|^{3l}$
vanish for $l>1$. 
The metric $h(\vec A)$ is harmonic for $d=3$ too though the moduli ${\vec A}$ do not exhaust 
all light degrees of freedom, the latter involving also the massless gauge field that is present in the effective Lagrangian.
(see Sect.\,3 for 
 more comments about the connection between the 4d nonrenormalization theorems and their lower 
dimensional counterparts.)

For the descendants of ${\cal N}\! =\! 1$ theories, multiloop corrections do not vanish.
Two-loop corrections were calculated in the Abelian case in 
 \cite{2loop}. For $d\! =\! 1$, the metric has the 
 form
 \be
\label{hN1}
h(\vec{A}) \ =\  1 + \frac {e^2}{2|\vec{A}|^3} - 
\frac {3e^4}{4|\vec{A}|^6} + \ldots \ ,\qquad {\vec A}\in{\mathbb R}^{3}\,.
  \ee  
This corresponds to massless SQED with one flavor.

The question arises whether these multiloop corrections are related to multiloop corrections in 4d 
$\beta$ 
function as they do for the  first loop.  
In Ref. \cite{2loop}, the calculations were performed
in component formalism and  this 
relationship  was not clearly seen. In this paper, we reproduce the calculations in the 
${\cal N} =1$ superfield formalism
and  establish such a relationship at the two--loop level.
We noted above that in $d\!=\!4$,  multiloop corrections to the $\beta$ function are related
to mass renormalization. One can ask  whether it is also true in some sense   in lower 
dimensions.
The answer to {\it this} question is negative.

In the main body of the paper we present a systematic study of one and two loop corrections
to the low energy effective Lagrangian of the Abelian and non-Abelian gauge theories for all 
dimensions $d\le 4$. At the perturbative level, it fixes the metric in the moduli space 
as well as dynamics of light degrees of freedom associated with the $d$ dimensional gauge fields
(present for $d=3,4$).

The plan of the paper is the following.  Sect.\,2 is the central part of the paper.
There we consider SQED and perform the superfield calculation of the 2--loop correction
 in different dimensions. 
In particular,   we demonstrate by diagrammatic methods how   the exact relation
(\ref{ephys1}) works  order by order in perturbation theory.
 In Sect.\,3, we discuss the ${\cal N}=2$ extension of SQED and emphasize 
the universal reason
by which the higher--order corrections to the metric vanish in all dimensions: 
extended supersymmetry requires it to be harmonic.   
We also illustrate how the two--loop corrections to the metric
cancel out: supergraph techniques allow one to  
  reproduce in a simple way the result of \cite{2loop} and extend it to all dimensions. 
In Sect.\,4, we extend the calculations to the non--Abelian case
emphasizing their relationship to Abelian ones. It also demonstrates the validity of the 
the known exact expression for the 4$d$ $\beta$ function \cite{NSVZ} at the two-loop  level.
The last section is reserved as usual to final conclusive remarks and acknowledgments. 

\section{Multiloop corrections to the metric: Abelian case}

\subsection{Notation and definitions}

 Let us start with fixing the notation.  
The density of Lagrangian of massive supersymmetric QED 
reads\,\footnote{\,Our conventions are close 
to that of Ref.\cite{WB},
$\theta^2 = \theta^\alpha \theta_\alpha$ ,\  $\int {\rm d}^2 \theta \, 
\theta^2 = \int {\rm d}^2 \bar\theta \, 
\bar\theta^2 = 1$; $(\sigma^\mu)_{\alpha \dot\beta} = 
\{1, \vecg{\sigma}\}_{\alpha \dot\beta}$,
$(\bar\sigma^\mu)^{\dot\beta \alpha} = \epsilon^{\dot\beta  \dot\gamma} \epsilon^{\alpha\delta}
(\sigma^\mu)_{\delta \dot\gamma} = 
\{1, -\vecg{\sigma}\}_{\dot\beta \alpha}$,
 but we use the metric
$\eta_{\mu\nu} = {\rm diag} (1,-1,-1,-1)$
 and include the extra
factor $2$ in the definition of $V$. Note that, for any superfield $X$, $\int d^2\theta X = 
- \frac {{\cal D}^2}4 X$ and $\int d^2\bar\theta X = 
- \frac {\bar{\cal D}^2}4 X$. }
 \be
\label{LSQED}
{\cal L} \ ={\rm Re}\left\{\frac{1}{2e^{2}}  \int {\rm d}^2\theta\, W^2 + 
   \int {\rm d}^4\theta \left[ \bar S \,e^{V} S+ 
\bar T \, e^{-V} T \right] + 
 2m\! \int \! {\rm d}^2\theta \, ST  \right\},
 \ee
where
\be
\label{V}
V &\!=&\!\! C + i\,\theta\chi - i\,\bar\theta \bar\chi + 
\frac i {\sqrt{2}} \,N\theta^2 -
\frac i {\sqrt{2}} \, \bar N \bar \theta^2 
 - 2 \theta \sigma^{\mu} \bar \theta A_\mu  \nonumber \\[1mm]
&\!+&\! \!\Big[ 2i\,\theta^2 \bar\theta \Big( \bar \lambda - 
\frac i4 \bar \sigma^{\mu} \partial_\mu   \chi \Big)
+ {\rm h.c.} \Big] + \theta^2 \bar \theta^2  \Big( D - 
\frac 14 \partial^2 C  \Big)
 \ee
and
 \be
 \label{Wdef}
  W_\alpha   
= \frac 18 \,\bar D^2 D_\alpha V  =
i(\lambda_\alpha + i\theta_\alpha D - \theta^\beta F_{\alpha\beta} - i \theta^2 \sigma^{\mu}
\partial_\mu \bar \lambda) \ 
 \ee
with 
   \be 
 \label{covdev}
&& \qquad {D}_\alpha = \frac {\partial}{\partial \theta^\alpha} - (\sigma^{\mu})_{\alpha \dot\alpha}
\bar \theta^{\dot\alpha} i\partial_\mu\,, 
\qquad   \bar{D}_{\dot\alpha} = - \frac {\partial}{\partial \bar\theta^{\dot\alpha}} + 
\theta^{\alpha} (\sigma^{\mu})_{\alpha \dot\alpha} i \partial_\mu\ ,
  \ee
  \be
\label{relD}
&&\{{D}_\alpha, \bar{D}_{\dot\beta}\} = 2i\left( \sigma^{\mu} \right)_{\alpha \dot\beta}
 \partial_\mu,\ \  \  {D}^2 \bar {D}^2 {D}^2 = - 16 \, \partial^2 {D}^2 , \ \  \
 \bar {D}^2 {D}^2  \bar {D}^2 = - 16 \, \partial^2  \bar {D}^2\ .
   \ee
The theory is defined by two parameters $e^{2}$ and $m$, which generically are  complex numbers.
Their imaginary parts lead to terms which are total derivatives in the Lagrangian density.
We can include such terms into consideration viewing 
$1/e^{2}$ and $m$ as spurion chiral fields.
In the 4d case one more scale parameter associated with the UV cut off should be added but 
for lower dimensions it is not needed.

The theory is invariant under gauge transformations,
\be
\label{gauge}
S\to {\rm e}^{-i\Lambda}S\,,\quad T\to {\rm e}^{i\Lambda}T\,,\quad V\to V+i(\Lambda -\bar\Lambda)
\ee
with the chiral function $\Lambda$. The superfield strength $W_{\alpha}$ is gauge invariant.
 We will also extensively use the superconnection $\Gamma_{\mu}$,
\be
 \label{Gammu}
\Gamma_\mu = \frac 14 (\bar \sigma_\mu)^{\dot \beta \alpha} \bar D_{\dot \beta} D_\alpha 
V = \ A_\mu +
\ldots \ ,
  \ee
which transforms as 
\be
\label{superg}
\Gamma_{\mu}\to \Gamma_{\mu} - \partial_{\mu}\Lambda\ .
\ee
The superconnection $\Gamma_\mu$ is defined such that the superfields $\nabla_\mu S =
(\partial_\mu - i\Gamma_\mu) S$ and  $\nabla_\mu T =
(\partial_\mu + i\Gamma_\mu) T$ are transformed under the gauge transformations (\ref{gauge}) 
in the same
way as $S,T$. Covariant derivatives acting on the right chiral superfields $\bar S, \bar T$
have the form $\partial_\mu \pm i \bar\Gamma_\mu$, where $\bar \Gamma_\mu$ is complex conjugate
of (\ref{Gammu}).

Note, that the superconnections $\Gamma_{\mu}$ are, of course, constrained superfields, 
it is the components of 
$V$ which are unconstrained.  The superderivatives of $\Gamma_{\mu}$ are related to the superfield 
strengths $W_{\alpha}$,  $\overline W_{\dot \alpha}$,
 \be
 \label{DGamma}
 \bar D_{\dot \alpha} \Gamma_{\mu}=(\sigma_{\mu})_{ \alpha \dot \alpha } W^{\alpha}\ ,
\qquad 
D_{\alpha} \overline\Gamma_{\mu}=(\sigma_{\mu})_{ \alpha \dot \alpha }{\overline W}^{\,\dot\alpha}
\ .
 \ee 
 ~\vspace{-5mm}
 
Our conventions for the supergraphs are close to those in Refs. \cite{Siegel,West}.
For convenience, we have included the factors $ {D^2}/4$ and 
$ {\bar D^2}/4$ 
usually attributed to the vertices in the definition of the propagators. In the absence of  
external field, the free 4-dimensional propagators  are
   \be
   \label{supprop}
\langle S_1 \bar S_2 \rangle \!&=&\! \langle T_1 \bar T_2 \rangle = 
\int\! \frac {{\rm d}^4 p}{(2\pi)^4}\, {\rm e}^{ip(x_2 - x_1)} \, \frac i{p^2 - |m|^2} \,  
\frac {\bar {D}^2_1 {D}^2_2 \delta^4(\theta_1 - \theta_2)}{16}\ ,
\nonumber \\[2mm]
\langle V_1 V_2 \rangle \!&=&\! -2e^2
\int\! \frac {{\rm d}^4 k}{(2\pi)^4}\, {\rm e}^{ik(x_2 - x_1)} \, \frac i{k^2} 
\left[1+(\alpha -1)\, \frac{{D}^2_1 {\bar D}^2_2 +\bar {D}^2_1 {D}^2_2}{16 k^{2}}\right]\delta^4(\theta_1 - \theta_2)
\ ,
          \ee
where  we should substitute $\pm p_\mu$  for $i\partial/\partial x_{1,2}^{\mu}$ in the expressions 
(\ref{covdev}) for ${D}_{\alpha}$ and    $\bar {D}_{\dot\alpha}$ .
Note the useful relation
   \be
 \label{D2exp}
\frac {\bar {D}^2_1 {D}^2_2 \delta^4(\theta_1 - \theta_2)}{16} \ =\ 
\exp \left\{p_\mu[\theta_2 \sigma_\mu \bar \theta_2 + \theta_1 \sigma_\mu \bar \theta_1 -
2 \theta_1 \sigma_\mu \bar \theta_2 ] \right\}\ .
  \ee
~\vspace{-5mm}

 The expression for the propagator of gauge superfield $V$ depends on 
the parameter $\alpha$ in the gauge fixing term 
\be
{\cal L}_{\rm g.f.}=- \frac{1}{32 e^2 \,\alpha} \int {\rm d}^4\theta \, D^2 V \bar D^2 V 
\ee
added to  the Lagrangian (\ref{LSQED}).    
We will use the Feynman gauge, $\alpha=1$. 
This gauge appears to be special providing for a benign behavior of the propagator
in the infrared. We will return to the discussion of this point later.
  
The vertices $\langle \bar S V^n S \rangle$ and $\langle \bar T V^n T \rangle$ 
are read out directly from the Lagrangian (\ref{LSQED}). 
Although the propagators $\langle \Phi_1 \Phi_2 \rangle$
and $\langle \bar \Phi_1 \bar \Phi_2 \rangle$ are also present in the theory, they involve 
the extra factors
$\bar \theta^2$ or $\theta^2$, and one can check that the contribution 
to ${\cal K}^{(2)}(\vecg{\Gamma})$ 
of the graph like in Fig.\ref{sgraph}b, but with the bars on the same line,  vanishes.

\subsection{Dimensional reduction}

The above definitions refer to SQED in 3+1 dimensions. To pass to the dimensionally reduced 
descendants of the theory, 
one can follow the standard logics (cf. e.g. \cite{Witten:df}) and put the system
in a small spatial box 
$|x_{k}|\le L_{k}/2$, $(k=1,2,3)$, imposing periodic boundary conditions. 
The gauge invariant integrals 
\be
I_{k}=\frac{1}{L_{k}}\int_{-L_{k}/2}^{L_{k}/2}A_{k}{\rm d}x_{k}
\ee
represent the moduli of the theory, they count the set of degenerate classical vacua. The moduli 
space is also a 
3-dimensional torus, $I_{k}$ are periodic coordinates living on the interval $|I_{k}|\le \pi/L_{k}$. 
A supersymmetric extension of $I_{k}$
\be
\tilde I_{k}=\frac{1}{L_{k}}\int_{-L_{k}/2}^{L_{k}/2}\Gamma_{k}{\rm d}x_{k}
\ee
invariant under supergauge transformations (\ref{superg}) adds fermionic moduli.

Reduction of the $k$-th spatial coordinate is introduced as the limit $L_{k}\to 0$. 
In this limit, $I_{k}$ coincides with $A_{k}$ (and $\tilde I_{k}$ with $\Gamma_{k}$)
while the interval where $A_{k}$ is defined becomes infinite.
For unreduced coordinates  we take an opposite limit $L_{k}\to \infty$ such that the corresponding 
moduli interval becomes a point $A_{k}=0$. Thus, we consider the limit when 
the moduli torus shrinks along unreduced directions and becomes unbounded along the reduced ones. 

The $d$-dimensional descendant of SQED represents a theory
with the Lagrangian density given by the same expression (\ref{LSQED}) but with the fields depending 
only on $d$ coordinates. The number of moduli fields is given by {\it co-dimension} $\hat d$ defined as
\be\label{co-d}
\hat d=4-d\ .
\ee
Thus, the moduli space of the theory has dimension $\hat d$ and is parametrized by the vector  
$\vec{A}=\{A_{\hat k}\}\in {\mathbb R}^{\hat d}$, where $\hat k$ marks reduced coordinates.  
We call this the Coulomb branch following a clear analogy with the ${\cal N}\!=\!2$ 
case \cite{SeiWit}, which can be viewed as a dimensional reduction from the ${\cal N}\!=\!1$ 
gauge theory in six dimensions. The coordinate dependent excitations $A_{\hat k}(x^{0},...,x^{d})$
 of moduli are called moduli fields.  The supersymmetric extension of these moduli fields 
is given by superfields $\Gamma_{\hat k}$ defined in Eq.\,(\ref{Gammu}).
It is important for us, however that the co-dimensional components
$\Gamma_{\hat k}$ (the supermoduli, $\hat k = d,\ldots,4$)
 are gauge invariant fields as it seen from \re{superg}. Moreover, $\Gamma_{\hat k}$ are real 
in contrast to the space--time $\bar \mu=0,\ldots, d\!-\!1$ components,  for which 
$\Gamma_{\bar \mu}-\overline\Gamma_{\bar \mu}
=i \partial_{\bar \mu} V$.\footnote{\,Gauge--invariant superconnections $\Gamma_{\hat k}$ were first introduced in \cite{scur}
in the context of UV regularization of 4d theories, where the limit
$\hat{d} \equiv \epsilon \to 0$ was studied.}

The set of light bosonic fields on the Coulomb branch is given by $A_{\mu}$. This includes, besides
gauge independent  moduli fields $\vec{A}$, the remaining gauge dependent $A_{\bar\mu}$ living 
in the $d$-dimensional space, $\bar \mu =0,\ldots, d-1$. 
We will see that  the 
effective Lagrangian depends not only on $W_\alpha$, but also on the supermoduli
$\Gamma_{\hat k}$ and, if limiting ourselves by the 
terms having not more than two derivatives of bosonic fields, can be written
in the form
   \be\label{genL}
{\cal L}_{\rm eff}={\rm Re}\left\{\frac{1}{2e^{2}}\!\int\! 
{\rm d}^2\theta\, h(\vecg \Gamma)\, W^{2} 
\right\},\qquad
\vecg{\Gamma}=\{ \Gamma_{1},..., \Gamma_{\hat d} \}\ .
    \ee
It differs from the original gauge term in \re{LSQED} by the function $h(\vecg \Gamma)$ 
which introduces dependence on gauge--invariant moduli superfields. 
Note that the  superfields $\vecg \Gamma$
are not chiral. However, it  is only the chiral part of $h(\vecg \Gamma)$ which contributes. 
Indeed, acting by $\bar D_{\dot \alpha}$ on the Lagrangian (\ref{genL}) we get zero as it follows 
from \re{DGamma} and $W_{\alpha}W_{\beta}W_{\gamma}=0$.

Trading the $\theta$ integration in \re{genL} for differentiation and using \re{DGamma} we 
obtain the component form of the effective Lagrangian,
\be
 \label{compL}
&&\hspace{-0.3cm}{\cal L}_{\rm eff} 
=
{\rm Re}\Bigg\{\frac{1}{2e^{2}}\bigg [h(\vec{A})\! 
 \bigg(\!\!-\!\frac 12 F_{\mu\nu}^{2}\!+\!D^{2}\!+\!2\lambda 
\sigma^{\mu}i\partial_{\mu}\bar\lambda
\! \bigg)\!-\!\frac i2  \frac{\partial h} {\partial A_{\!\hat k}} \, F_{\mu\nu} 
\lambda \sigma^{\mu}\bar\sigma^{\nu}\sigma_{\!\hat k}\bar\lambda - 
\frac{\partial h} {\partial A_{\!\hat k}} \, D 
\lambda \sigma_{\!\hat k}\bar\lambda
\nonumber\\[1mm]
&&\hspace{2.7cm} - \frac 14 \, \frac {\partial^{2} h}{\partial  A_{\!\hat k}\partial  A_{\!\hat k}} 
\, \lambda^{2}{\bar \lambda}^{2}\bigg]\Bigg\}\,.
 \ee
 We have omitted here the terms which represent 
the total derivatives, like $\theta$-term.
The part of the Lagrangian (\ref{compL}) which contains bosonic moduli fields $A_{\hat k}$,
 \be
\label{bosmod}
 {\cal L}_{\rm eff}^{\rm moduli}=\frac{h(\vec A)}{2e^{2}}\partial_{\mu}{\vec A}_{\hat  k}\, 
\partial^{\mu}\!{\vec A}_{\hat  k} \ ,
\ee
shows that the metric in the moduli space has a simple conformally flat form,
\be
{\rm d}s^{2}=h^{\hat k \hat l}(\vec A){\rm d}A_{\hat k}\,{\rm d}A_{\hat l}\,,\qquad
h^{\hat k\hat l}(\vec A)=h({\vec A})\,\delta^{\hat k \hat l}\ .
\ee

We also can present  ${\cal L}_{\rm eff}$ as an integral over full superspace,
\be
\label{Leffd}
{\cal L}_{\rm eff}= {\rm Re}\Bigg\{\frac{1}{e^{2}}\!\int\! {\rm d}^4\theta\,   {\cal K} 
({\Gamma})\Bigg\} \ , 
 \ee 
where the function ${\cal K} ({A})$ is related to the metric $h(\vec{A})$ via derivatives,
\be
\label{Kh}
h(\vec A)\ =\  
\frac 12\, \frac{\partial^{2} {\cal K}(A)}{\partial A^{\mu}\,\partial A_{\mu}} \ . 
 \ee
The equivalence of (\ref{genL}) and (\ref{Leffd}) is revealed when substituting 
$\int d^2\bar \theta$
by $- \bar {\cal D}^2/4$ and using the relations   (\ref{DGamma}).
Note that ${\cal K} $ depends on all four components of $A_{\mu}$ while the metric $h$ is the 
function of 
$\hat d$-dimensional moduli only.  Equation  (\ref{Kh}) implies that we can consider 
${\cal K} $  as a function of the moduli $\vec A$ up to harmonic terms such as 
\be
\label{harm}
\frac{1}{d}\,A_{\bar \mu}A^{\bar \mu}+\frac{1}{\hat d}\, \vec A^{2}\ . 
\ee
This harmonic ambiguity in ${\cal K}$ is  particularly important in the limit of four dimensions 
when $\hat d \to 0$.

We can illustrate this  in the leading classical approximation.
In this approximation the metric is
\be
\label{treeh}
 h^{(0)}(\vec A)=1\ ,
\ee
and the function $K$ satisfying \re{Kh} and depending only on $\vec{A}$ is
\be
\label{treeK}
{\cal K}^{(0)}(\vec A)=-\frac{1}{\hat d }\,{\vec A}^{2}\ ,
\ee
which is singular at $\hat d \to 0$. Adding the harmonic term (\ref{harm}),  we come to
\be
\label{treeK1}
{\cal K}^{(0)}( A)=-\frac{1}{d }\,A_{\bar \mu}A^{\bar \mu}\ ,
\ee
which allows for a smooth $\hat d \to 0$ limit.  Although, in contrast to \re{treeK}, this form of  $K$ 
is not gauge 
invariant  both forms lead to the same gauge invariant action.

\subsection{One-loop corrections}

The one-loop corrections  due to the charged 
matter fields, Fig.\,\ref{sgraph}a, were calculated in Ref. \cite{LeffN1,Akhmedov}.
For any dimension $d$ in the interval $1\le d \le 4$, the one-loop expression for $h^{(1)}(\vec A)$ 
can be presented in the form of the Feynman integral (after the Wick rotation of $p_{0}$)
\be
\label{h1}
h^{(1)}(\vec A)=2e^{2}\int \frac{{\rm d}^{d} p}{(2\pi)^{d}}\left[\frac{1}
{(p^{2}+\vec A^{2 }+|m|^{2})^{2}}
-\frac{1}{(p^{2}+\vec A^{2}+ \Lambda_{\rm UV}^{2})^{2}}\right],
\ee
where $A_{\hat k}$  is substituted for co-dimensional components of the momentum $p_{\hat k}$ and 
the second 
term in the square brackets introduces Pauli-Villars regularization needed only for $d=4$.
Generically, $\Lambda_{\rm UV}$ might be complex, but we assume it to be real and positive.
Probably, the simplest way to find $h$ is  to calculate the loop in the
 background 
of constant potential $A_{\mu}$ and constant auxiliary field $D$. 
Then $A_{\mu}$ enters as a shift
$p_{\mu}\to p_{\mu}+A_{\mu}$ while $D$ splits fermion and boson masses. 
Expanding over $D$ up to the second order and comparing the result with the
corresponding term in Eq.\,(\ref{compL}), one arrives at (\ref{h1}).
The result of integration is
\be
\label{1loop}
h^{(1)}(\vec A)=\frac{2\Gamma({\hat d}/2)}{(4\pi)^{d/2}}\,\frac{e^{2}}
{({\vec A}^{2}+|m|^{2})^{{\hat d}/2}}- \bigg(m\to\Lambda_{\rm UV}\bigg)\ .
\ee

Note that the integral (\ref{h1}) is convergent in the infrared and the result (\ref{1loop}) 
implies the choice of low normalization point, $\mu \ll (|m|^{2}+{\vec A}^{2})^{1/2}$, for 
the effective Lagrangian.
 Let us add also that, once $h(\vec A)$ is known,  the function ${\cal K}(\vec A)$ is defined 
 by integration. We will give later explicit expressions for ${\cal K}(\vec A)$ for
different integer $d$.

\subsection{Two-loop calculations}

At the two--loop level, only the graph depicted in
Fig.\,\ref{sgraph}b (or rather two such graphs with chiral superfields $S,T$ in the loop) 
contribute.
Note that the graphs involving 4-point vertices $ \bar \Phi \Phi vv $,  
 do not contribute because in the Feynman gauge we are using the quantum superfield propagator
$\langle v_1 v_2 \rangle \propto\delta^4 (\theta_1 - \theta_2)$, which vanishes
at the coinciding superspace points. 
Comparing the general form (\ref{Leffd}) of the effective Lagrangian with the expression 
corresponding to the graph in Fig.\,\ref{sgraph}b, 
we get the expression for the two-loop part of ${\cal K}$,
 \be
{\cal K}^{(2)}\left(\vecg\Gamma(z_{1})\right)= i e^{2}\! \int {\rm d}^{d}x_2 {\rm d}^4\theta_2 
\langle S_1 \bar S_2 \rangle  \langle  S_2 \bar S_1  \rangle
\langle v_1 v_2 \rangle \ .
  \ee
  
Generically, the expression for the superpropagator $\langle S_1 \bar S_2 \rangle $ in external
gauge field is rather complicated. It has the form
  \be
\langle S_{1} \bar S_{2} \rangle= -\frac{i}{16}\Big[\nabla_\mu \nabla^\mu \!+\! 
iW^\alpha \nabla_\alpha \!+ \!
\frac i2 (\nabla^\alpha W_\alpha) \!+\! |m|^2 \Big]^{\!-1}
\bar {\nabla}_{1}^{2}{\nabla}_{2}^{2}\,\delta^{d}(x_{1}-x_{2}) \delta^{4}(\theta_{1}-\theta_{2})\,,
 \ee
 where all derivatives are covariant, see Ref. \cite{Siegel} for details.

However, we can use {\it any} background for determination of ${\cal K}^{(2)}(\vecg\Gamma)$.
A convenient choice is to keep only the
lowest component $\Gamma_k|_{\bar \theta = \theta = 0} = A_k$ and assume that all higher components
 vanish.\footnote{\,The effective action $\int {\rm d}^4\theta \, {\cal K}^{(2)} $ 
vanishes with this choice, but
the {\it integrand} ${\cal K}^{(2)}$ does not. Having found the function 
${\cal K}^{(2)}(\vecg{\Gamma})$, we are 
going to
use it for an arbitrary background with nonvanishing $W_\alpha$ giving a nontrivial contribution in the
effective action.}
For such a choice, 
the superfield $W_\alpha$ vanishes and the propagator of the
charged superfields has a particular simple form. In momentum space it can be obtained from 
the free 4d propagator
(see the footnote at the previous page) by 
substituting moduli $A_{\hat k}$ for the co-dimensional components of momentum ${p}_{\hat k}$. 
   \be
\label{S1S2}
\langle S_1 \bar S_2 \rangle = \int \frac {{\rm d}^{d} p}{(2\pi)^{d}}\, {\rm e}^{i p(x_1-x_2)} 
\frac i {p^2-\vec{A}^2 -|m|^{2}} \, 
\frac {\bar {D}^2_1 {D}^2_2 \delta^4(\theta_1 - \theta_2)}{16} \ .
   \ee
Bearing all this in mind and using the relation (\ref{D2exp}) and its corollary\,\footnote{\,Note that one needs at least 4 spinor derivatives to deal with 
the second $\delta$ function
and to obtain a nonvanishing result. This was exactly the reason by which we were allowed
to substitute the covariant derivative $\nabla_\alpha$
by  $D_\alpha$ in the propagator
$\langle S_1 \bar S_2 \rangle$.} 
 \be
 \label{reldel}
 \delta^4(\theta_1 - \theta_2) \, \frac {\bar {D}^2_1 {D}^2_2} {16}   \, 
\delta^4 (\theta_1  - \theta_2) \ = 
\  \delta^4 (\theta_1  - \theta_2)\ ,
  \ee 
we can write after Wick rotation
  \be
\label{KGamres}
 {\cal K}^{(2)}(\vec{A}) \ =\ -2 e^4  \int \frac {{\rm d}^{d}k}{(2\pi)^{d}}\,\frac{1}{k^2}
\int \frac {{\rm d}^{d} p}{(2\pi)^{d}} \, \frac 1{p^2 + \vec{A}^2+|m|^{2}} \,
\frac 1{(p+k)^2 + \vec{A}^2+|m|^{2}}\ .
  \ee
In this expression $p$ and $k$ refer only to the $d$-dimensional part of momenta, the 
co-dimensional part of $p^{2}$ is written explicitly via ${\vec A}$. 

Let us start the calculation with representing the integral over $p$ in the dispersion form,
\be
\label{disp}
&& \int \frac {{\rm d}^{d} p}{(2\pi)^{d}} \, \frac 1{p^2 + \vec{A}^2+|m|^{2}} \,
\frac 1{(p+k)^2 + \vec{A}^2+|m|^{2}}=\int_{4({\vec{A}}^2+|m|^{2})}^{\infty}\!{\rm d}s\,
\frac{\rho(s,m,\vec{A})}{s+k^{2}}\ ,
\nonumber\\[2mm]
&& \rho(s,m,\vec{A})=\frac{1}{\sqrt{s}\, (s-4\vec{A}^2-4|m|^{2})^{(\hat d-1)/2}}\, 
\frac{1}{2^{2d-3}\pi^{(d-1)/2}\Gamma(\frac{3-\hat d}{2})}\ .
  \ee
To make the integral over $s$ convergent at $d=4$, we
regularize it in the UV by the Pauli-Villars regulators, i.e., by 
subtracting from the integrand the expression where $m$ is substituted by $\Lambda_{\rm UV}$,
\be
\rho_{\rm reg}(s,m,\vec{A})=\rho(s,m,\vec{A})-\rho(s,\Lambda_{\rm UV},\vec{A})\ .
\ee
The integral over $k$ in (\ref{KGamres}) involves for $d\le 2$  an infrared divergence 
coming from the small $k$ region. It also needs the UV regularization for $d=4$.
By substituting the photon propagator 
  \be
 \label{infrmu1}
\frac{1}{k^{2}}\to  \frac{1} {k^{2}+\mu^{2}}- \frac{1} {k^{2}+ \Lambda_{\rm UV}^{2}}\ .
  \ee
we take care of both IR and UV divergences. (We have chosen the ultraviolet regulator
for the photon propagator to be the same as the Pauli--Villars regulator, though in principle
they could be different.) Integrating then over $k$, we get 
\be
\label{sint}
&&{\cal K}^{(2)}\!(\vec{A})=e^{4}\int_{4(\small{\vec{A}}^2+|m|^{2})}^{\infty}\!{\rm d}s\,
\frac{\rho_{\rm reg}(s,m,\vec{A})}{2^{d-1}\pi^{(d-2)/2}\Gamma(\frac{4-\hat d}{2})
\sin\frac{\hat d -2}{2}} \, 
\frac{s^{(2-\hat d)/2}-(\mu^{2})^{(2-\hat d)/2}}{s-\mu^{2}} 
\nonumber\\[2mm]
&& \qquad\qquad\;-\left(\mu^{2}\to \Lambda_{\rm UV}^{2}\right)
\ .
\ee
We need the piece where $\mu^{2}$ is substituted by $\Lambda_{\rm UV}^{2}$ only for $d=4$.
In this case the  $\Lambda_{\rm UV}^{2}$ part cancels completely the part from the first line in 
\re{sint}
and both ${\cal K}^{(2)}$ and $h^{(2)}$ vanish at $d=4$. We will  discuss this vanishing in 
more detail in the next Section.

For $d<4$ we obtain
\be
\label{K2d}
&&{\cal K}^{(2)}\!(\vec{A}) =  \frac{\sqrt{\pi}\Gamma(\frac{\hat d}{2}\!-\!1)
\Gamma(\hat d \!-\!1)}{4(2\pi)^{d}\Gamma(\frac{\hat d\!+\!1}{2})}\, \frac{e^{4}}
{(\vec{A}^2\!+\!|m|^{2})^{\hat d \!-\!1}}
-\frac{2\Gamma(\frac{\hat d}{2}\!-\!1)\Gamma(\frac{\hat d}{2})}{(4\pi)^{d}}\,
\frac{e^{4}}{\mu^{\hat d\! -\!2}(\vec{A}^2\!+\!|m|^{2})^{\hat d/2}}
\nonumber\\[2mm]
&& \qquad\qquad\;-\left(m\to \Lambda_{\rm UV}\right)
\ ,
 \ee
 and the corresponding metric $h^{(2)}$ is
 \be
\label{h2d}
&& h^{(2)}\!(\!\vec{A})\! =\!  -\frac{\sqrt{\pi}\Gamma(\frac{\hat d}{2}\!-\!1)
\Gamma(\hat d \!+\!1)}{4(2\pi)^{d}\Gamma(\frac{\hat d\!+\!1}{2})}\, \frac{e^{4}
({\vec A}^{2}\!-\!|m|^{2})}{(\vec{A}^2\!+\!|m|^{2})^{\hat d \!+\!1}}
+\frac{4\Gamma(\frac{\hat d}{2}\!-\!1)\Gamma(\frac{\hat d}{2}\! +\! 1)}{(4\pi)^{d}}\,
\frac{e^{4}(\vec{A}^2\!-\!\frac{\hat d}{2}|m|^{2})}{\mu^{\hat d\! -\!2}
(\vec{A}^2\!+\!|m|^{2})^{2\!+\!\hat d/2}}
\nonumber\\[2mm]
&& \qquad\qquad\;-\left(m\to \Lambda_{\rm UV}\right)
\,.
 \ee
The apparent singularities at $\hat d = 1,2$ cancel out as they should. Indeed, introducing
the infrared
and ultraviolet regularizations as in Eq.\,(\ref{infrmu1}) makes the integral (\ref{KGamres}) 
finite for any $d$.

 The singular in the infrared parameter $\mu$ part of this result coincides
 with the one produced by the one-loop correction to matrix element of  ${\cal K}^{(1)}$,  
see \cite{2loop} for detailed discussion. Indeed, denoting $a_{k}$ deviations in the moduli 
$A_{k}$, we get 
for this matrix element
 \be
 \label{meK1}
 \langle {\cal K}^{(1)}\rangle_{\rm 1-loop}&\! =\! &  
\frac 12\,\frac{\partial^{2}{\cal K}^{(1)}}{\partial A_{k}\partial A_{l}} \langle a_{k} 
a_{l}\rangle= \frac{e^{2}}{2}\,
\frac {\partial^2 {\cal K}^{(1)}} {\partial A_{k}\partial A_{k}}  
\int \! \frac 
{{\rm d}^{d}k}{(2\pi)^{d}}\left[\frac{1}{k^2 +\mu^2 } -\frac{1}{k^2 +\Lambda_{\rm UV}^2 }\right]
\nonumber\\[1mm]
&\! =\! &-\frac{e^{2}\, h^{(1)}\!(\vec A)\Gamma(\frac{\hat d}{2}\!-\!1)}{(4\pi)^{d/2} 
\mu^{\hat d\!-\!2} }
 -(\mu \to \Lambda_{\rm UV})\ ,
     \ee 
The matching of infrared-singular parts in expressions (\ref{K2d}) and (\ref{meK1}) 
means that, 
when we pass to the Wilsonean effective action, the IR parameter $\mu$ 
 in Eqs.\,(\ref{K2d}) and (\ref{h2d}) becomes the
normalization point  for the Wilsonean action. We implied in the above expressions that the 
normalization point $\mu$ is much less than $(\vec{A}^2\!+\!|m|^{2})^{1/2}$.

\subsubsection{Dimension 1}

In SQED reduced to one dimension,
 the moduli space is three-dimensional with the following metric $h$:
\be
h_{1d}(\vec{A})=1+\frac{e^{2}}{2\,(\vec{A}^2\!+\!|m|^{2})^{3/2}}- 
\frac {3e^4({\vec A}^{2}-|m|^{2})} {4({\vec A}^{2}+|m|^{2})^4}+\frac 
{3e^4(2{\vec A}^{2}-3|m|^{2})} {16\,\mu\,({\vec A}^{2}+|m|^{2})^{7/2}}\ .
\ee
Here the tree, \re{treeh}, one-loop, \re{1loop} and two-loop, \re{h2d}, terms are combined.
The effective Lagrangian is given by \re{compL} where the only nonvanishing components of 
$F_{\mu\nu}$ are $F_{0k}=\dot A_{k}$. 

The function ${\cal K}(\vecg \Gamma)$ is
\be
{\cal K}_{1d}(\vecg \Gamma)=-\frac 13\, {\vecg\Gamma}^{2}
+\frac{e^{2}{\rm arcsinh}({\vecg \Gamma}^{2}/|m|^{2})^{1/2}}{({\vecg \Gamma}^{2})^{1/2}}
+\frac {e^4} {8({\vecg \Gamma}^{2}+|m|^{2})^2}-\frac {e^4} {4\,\mu\,
({\vecg \Gamma}^{2}+|m|^{2})^{3/2}}\ .
\ee

\subsubsection{Dimension 2}

The moduli space in the SQED reduced to two dimensions $x^{0}, x^{1}$ is parametrized by 
$\vec{A}=\{A_{2},A_{3}\}$ with the metric
\be
\label{metrd2}
h_{2d}(\vec{A})=1+\frac{e^{2}}{2\pi}\,\frac{1}{\vec{A}^2\!+\!|m|^{2}}
+\frac{e^{4}}{4\pi^{2}}\left[\frac{\vec{A}^2\!-\!|m|^{2}}{(\vec{A}^2\!+\!|m|^{2})^{3}}
\ln\frac{\vec{A}^2\!+\!|m|^{2}}{\mu^{2}}-\frac{|m|^{2}}{(\vec{A}^2\!+\!|m|^{2})^{3}}
\right],
\ee
as it follows again from Eqs.\,(\ref{treeh}), (\ref{1loop}) and the $d\to 2$ limit 
of \re{h2d}.\footnote{\,Actually, the parameter $\mu$ in Eq.\,(\ref{metrd2}) differs from 
that in Eq.\,(\ref{h2d})
by a certain constant factor chosen such that the two--loop term in Eq.\,(\ref{metrd2}) is a pure
logarithm in the limit $m=0$. }
The effective Lagrangian (\ref{compL}) contains besides moduli fields $\vec{A}=\{A_{2},A_{3}\}$ 
the vector field $A_{\bar \mu}(x^{0}, x^{1})$ ($\bar \mu=0,1$). Although $A_{\bar \mu}$ contains 
no propagating degrees of freedom, it induces a contact interaction for fermionic fields, like 
 the auxiliary field $D$ does.
The function ${\cal K}(\vecg \Gamma)$ in this case has the form
\be
\label{2dK2}
{\cal K}_{2d}(\vecg \Gamma)=-\frac 12\, {\vecg\Gamma}^{2}-\frac{e^{2}}{4\pi}
\int_{0}^{\frac{\vec{\small\Gamma}^{2}}{|m|^{2}}}\!\!\!{\rm d}t\,\frac{\ln(t+1)}{t}
-\frac{e^{4}}{8\pi^{2}}\,\frac{1}{\vecg{\Gamma}^2\!+
\!|m|^{2}}\left[\ln\frac{\vecg{\Gamma}^2\!+\!|m|^{2}}{\mu^{2}}+2\right].
\ee

Now, the bosonic part of the effective Lagrangian has the sigma model form (\ref{bosmod}). 
The full effective Lagrangian represents its supersymmetric extension with two complex 
supercharges. A theorem due to Alvar\'ez-Gaume and Freedman \cite{AGF} says that there 
is only {\it one} such Lagrangian   representing
a K\"ahler supersymmetric $\sigma$ model (incidentally, 
any 2-dimensional manifold is K\"ahlerian).  
In other words, the effective Lagrangian can be presented in the form
 \be
 \label{d2Kal}
{\cal L}_{\rm eff} \ =\ \frac 1{e^2} \int {\rm d}^4\theta \, {K}(\bar\Phi, \Phi)\ ,
\ee
where 
$
\Phi\!  =\! \phi\! + \sqrt{2}\theta \lambda\!  +\theta^{2}\! F
$
 is a chiral superfield  with the lowest component $\phi\!  = \! (A_2 \! +\!  i A_3)/\! \sqrt{2}$. 
This was emphasized in Ref.\cite{Akhmedov}, where the K\"ahler potential ${K}(\bar\Phi, \Phi)$
was calculated at the one-loop level. 
One can explicitly check that 
the component form of the Lagrangian (\ref{d2Kal}) coincides, indeed, with (\ref{compL})
if the K\"ahler  potential $K$ is chosen as 
\be
{K}(\bar\Phi, \Phi)\ =\ -{\cal K}_{2d}({\vecg \Gamma}^{2}\!=\!2\bar\Phi \Phi)\ .
\ee

\subsubsection{Dimension 3}

Let us denote $A$ the only modulus that survives at $d\!=\!3$.  The metric becomes
\be
h_{3d}(A)=1+\frac{e^{2}}{4\pi}\,\frac{1}{({A}^2\!+\!|m|^{2})^{1/2}}+
\frac{e^{4}}{16\pi^{2}}\, \frac{{A}^2\!-\!|m|^{2}}{({A}^2\!+\!|m|^{2})^{2}}\ ,
\ee
and the function ${\cal K}(\Gamma)$ is
\be
\label{K3d}
{\cal K}_{3d}(\Gamma)=-\Gamma^{2}-\frac{e^{2}}{2\pi}\left[\Gamma\,{\rm arcsinh}\,\frac{\Gamma}{|m|} 
- \sqrt{\Gamma^{2} +|m|^{2}}\right] +\frac{e^{4}}{16\pi^{2}}\ln\frac{\Gamma^{2} +|m|^{2}}
{\Lambda_{\rm UV}^{2}}\ .
\ee
If the field $A$ were the only bosonic field in the effective theory, the metric could be made 
trivial by field redefinition. However, at $d\!=\!3$, unlike what happens at
 $d\!=\!1,2$, the vector field 
$A_{\mu}$
describes one propagating degree of freedom. 

The effective Lagrangian (\ref{compL}) takes the following 3d form:
\be
{\cal L}_{\rm eff}\! =\! \frac{1}{2e^{2}}\! \left\{\!h\bigg[\!-\! \tilde F_{\mu}^{2}\!+
\!(\partial_{\mu}A)^{2}\!+\! (\lambda \sigma^{\mu}i\partial_{\mu}\bar\lambda\!+\!{\rm h.c.})\! 
\bigg]\! -
h' \tilde F_{\mu}\lambda \sigma^{\mu}\bar\lambda
-\frac 14\bigg(h''\!-\!\frac{(h')^{2}}{2h}\bigg)\lambda^{2}\bar\lambda^{2}\!
\right\}\! .
\ee
Here $\tilde F_{\mu}=\epsilon_{\mu\nu\gamma}F^{\nu\gamma}/2$ and we excluded the auxiliary field 
$D$. Let us now introduce the dual photon field $\pi$ performing the duality transformation,
i.e.  adding the term $\partial^{\mu} \pi\tilde F_{\mu}/e^{2}$ to the Lagrangian and integrating 
over $\tilde F_{\mu}$. 
We obtain
\be
\label{duaL}
{\cal L}_{\rm eff}=\frac{1}{2e^{2}}\left\{\frac{1}{h}\bigg[(h\partial_{\mu}A)^{2}+
(\partial_{\mu}\pi)^{2}\bigg]
+h(\lambda \sigma^{\mu}i\partial_{\mu}\bar\lambda\!+\!{\rm h.c.})
-\frac 14\bigg(h''\!+\!\frac{(h')^{2}}{h}\bigg)\lambda^{2}\bar\lambda^{2}\!
\right\}.
\ee

To compare this expression with the generic form (\ref{d2Kal}) of K\"ahler model, 
let us relate our fields with components  of the chiral superfield $\Phi$ in the following way:
\be
\Phi =\phi +\sqrt{2}\, \theta \psi +\theta^{2} F\,, 
\quad \phi= \frac{\sigma+i\,\pi}{\sqrt{2}}\,,\quad \sigma=-\frac 12 \,{\cal K}'(A)\,\quad
\quad \psi=h(A)\lambda\,.
\ee
We verify then that the effective Lagrangian (\ref{duaL})
can be presented in the  K\"ahler form (\ref{d2Kal})  with the  K\"ahler function 
$K(\frac{\Phi+\bar\Phi}{\sqrt{2}})$
depending only on the sum $\Phi +\bar\Phi$ and satisfying the relation
\be
\frac 12 \,K''(\sigma)\!=\!
\frac{1}{h[A(\sigma)]}\!=\!
1\!-\!\frac{e^{2}}{4\pi({\sigma}^2\!+\!|m|^{2})^{1/2}}\!-\!
\frac{e^{4}}{16\pi^{2}}\!\bigg[ \frac{\sigma{\rm arcsinh}(\sigma/|m|)}{({\sigma}^2\!+
\!|m|^{2})^{3/2}}
\!-\!\frac{2|m|^{2}}{({\sigma}^2\!+\!|m|^{2})^{2}}\bigg],
\ee
where $\sigma=(\phi+\bar\phi)/\sqrt{2}=-{\cal K}'(A)/2$.

Thus, we came to the K\"ahler model where metric does not depend on the extra modulus 
$\pi$ associated with the dual photon. It means that $\pi$ is a coordinate along an isometry 
of the metric.
The phase nature of  $\pi$ becomes  more visible if  one would use the superfield 
$\tilde \Phi=\mu\exp(\Phi/\mu)$
instead of $\Phi$ (with $\mu$ being an arbitrary scale parameter). The metric depends only 
on ${\bar {\tilde \Phi}} \tilde \Phi$.

\subsection[$d=4$: Anatomy of zero]{{\boldmath $d=4$}: Anatomy of zero}

As was discussed in the Introduction, in massive 4d SQED, the 2--loop contribution to ${\cal L}_{\rm eff}$
vanishes. We have seen 
before how this comes about when explicitly evaluating 
 the expression (\ref{KGamres}) for ${\cal K}^{(2)}$ carefully regularized in the ultraviolet and
infrared [see the remark after Eq.\,(\ref{sint})]. The zero was obtained after cancellation of infrared
and ultraviolet contributions to the integral. 
Indeed, consider the expression (\ref{h2d}) for the metric at $d < 4$. It was obtained from the
first term in Eq.(\ref{sint}) for $\mu \ll m$. 
The second term corresponding to the UV regularization
of the photon propagator does not contribute at $d<4$, but for $d=4$ it is important. 
In the limit $4-d = \hat{d} \equiv \epsilon \to 0$, Eq.(\ref{h2d}) gives
  \be
h^{(2)}=-\frac{e^{4}}{32 \pi^{4}}\log\frac{\Lambda_{\rm UV}^{2}}{|m|^{2}}\ .
\ee
This is canceled by the UV contribution coming from the second term in
Eq.(\ref{infrmu1}).

It is instructive to explore in more details the mechanism for this cancellation in four
dimensions redoing the calculations 
 in a different and more transparent way.
Note first of all that there are  
no moduli in $d=4$ and the integral (\ref{KGamres}) gives just a number -- 
not a function of connections. On the other hand, in 4 dimensions the correction (if any)  is 
proportional to the original Lagrangian of the gauge field
 \be
 \label{WGam}
\frac{C}{2e_{0}^{2}}\int {\rm d}^2\theta \, W^2  \ =\  \frac{C}{4e_{0}^{2}} \int {\rm d}^4\theta \, 
\Gamma_\mu^2\ .
 \ee
Hence,  $C=0$.

Let us forget about this vanishing for the moment and try to calculate the contribution
in the same way as we did it for lower dimensions. 
As earlier, we consider 
the background involving only the component $A_\mu$ that is constant. 
The expression for ${\cal K}^{(2)}$ can be written in the form (before    
Euclidean rotation),
  \be
\label{KGamres4}
 {\cal K}^{(2)} = 
 -2 e_0^4  \int \frac {{\rm d}^4k}{(2\pi)^4 k^2}
\int \frac {{\rm d}^4p}{(2\pi)^4} \ \frac 1{(p_\mu  +{\Gamma}_\mu)^2 - m^2} \,
\frac 1{(p_\mu  + k_\mu + {\Gamma}_\mu)^2 - m^2}\ .
  \ee
With $\Gamma_\mu$ representing just a shift of the variable of integration, the integral 
obviously does not depend on $\Gamma_\mu$. This notwithstanding, expand the right side
of Eq.\,(\ref{KGamres4}) in $\Gamma_\mu$ and keep the quadratic terms.  Using Lorentz symmetry and 
performing Wick rotations in the momentum integrals, we obtain
  \be
  \label{Kbox}
 {\cal K}^{(2)}  \ =\ \frac{e_0^4}4\, \Gamma_\nu^2   \int \frac {{\rm d}^4k}{(2\pi)^4 k^2}
\int \frac {{\rm d}^4p}{(2\pi)^4} \, \frac{\partial^{2}}{\partial p_{\mu}^{2}}\left[ 
\frac 1{p^2 + m^2}\, 
\frac 1{(p  + k)^2 + m^2}\right] \ .
 \ee
The integral is still zero. The integrand can be represented as 
  \be
 \label{box2}
 \frac{\partial^{2}}{\partial p_{\mu}^{2}}\left[ \frac 1{p^2 + m^2}\, 
\frac 1{(p  + k)^2 + m^2}\right] 
\ =\ - \frac {8m^2}{(p^2 + m^2)^3}\, 
\frac 1{(p  + k)^2 + m^2}  \nonumber \\
 -  \frac 1{p^2 + m^2}\, \frac {8m^2}{[(p+k)^2 + m^2]^3} 
+ \ \frac{8(p^2 + pk)}{(p^2+m^2)^2 [(p+k)^2 + m^2]^2} \ .
 \ee
Now, the first two terms in the right side are infrared in nature. Indeed, the integral 
over the first term 
is saturated by $p^{2}\sim m^{2}$. In the limit $ m \to 0$,  it can be assimilated
 to the integral of $\delta$ function,
$$
- \frac  {8m^2}{(p^2 + m^2)^3} \ \to \ -4\pi^2 \delta^{4}(p)\ .
$$
Similarly,  the second term goes over to $\delta^{4}(p+k)$ in the massless limit.
On the other hand, the third term is not infrared. The corresponding
integral is saturated in the region $p^2 \sim k^2$ and integration over $k$ is logarithmic, i.e.,
$m^2 \ll k^2 \ll \Lambda_{\rm UV}^2$. 

Everything is prepared now to determine the effective Lagrangian at the normalization point 
$\mu \gg m$ or, in other terms,  in {\it
massless} theory. Indeed, the  effective Lagrangian is defined after
integrating over all modes with characteristic energy exceeding the
separation scale $\mu$. Up to now, we assumed that $\mu \ll m$. But in the opposite
limit, the infrared contributions in the integral coming from the virtualities
$p^2 \sim m^2$ should simply be discarded because this range of momenta should not be counted.
These momenta will reappear in the matrix elements in the effective theory.
Only the third term in R.H.S. of Eq.\,(\ref{box2}) is left. It gives a nonvanishing
contribution to the integral.

The latter is conveniently evaluated as the infrared contribution taken with opposite sign.
We obtain
  \be
\label{K4res}
{\cal K}^{(2)}(\mu) \!\ = \!\  2e_0^4 \pi^2 \, \Gamma_\mu^2  \! \int \!\!\frac {{\rm d}^4k}{(2\pi)^4 k^2}\!
\int \!\!\frac {{\rm d}^4p}{(2\pi)^4}  \, \frac{\delta^4(p)}{(p+k)^2} =
\frac {e_0^4 \pi^2}{2(2\pi)^6} \, \Gamma_\mu^2 \!  \int_\mu^{\Lambda_{\rm UV}}\!\!  \frac {d^4k}{ k^4}
= \frac {e_0^4\,\Gamma_\mu^2}{64\pi^4} \ln \frac {\Lambda_{\rm UV}}{ \mu} \, .~
  \ee
Bearing in mind the relation (\ref{WGam}) and the definitions (\ref{LSQED}), (\ref{Leffd}), 
we obtain the contribution 
$(e_0^2/16\pi^4) \ln (\Lambda_{\rm UV}/\mu) $ in $1/e^2(\mu)$ 
in accordance with Eq.\,(\ref{ephys}). 

\begin{figure}[h]
   \begin{center}
 \includegraphics[width=4.8in]{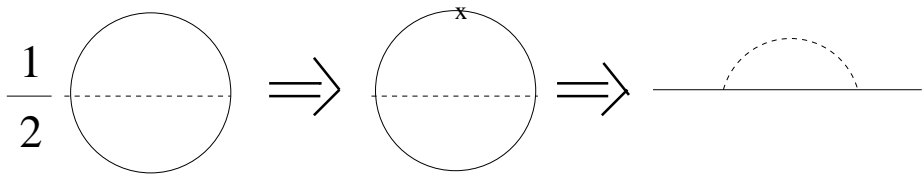}
        \vspace{-2mm}
    \end{center}
\caption{\small Two--loop effective action via one--loop polarization operator.}
\label{2petIR}
\end{figure}

By construction, the 2--loop contribution to $1/e^2(\mu)$ that we have just evaluated depends on the
graph in Fig.\,\ref{sgraph}b with very small virtuality of one of the matter field lines, i.e. this
line is effectively cut off. The diagram thus obtained describe the one--loop polarization operator 
of the matter field $\Sigma^{(1)}(p)$. To make things absolutely clear, we illustrate the 
procedure just described in Fig.\,\ref{2petIR}.\,\footnote{\,A 
factorization of similar nature was observed in ${\cal N}=4$ SYM for more complicated case of  scattering amplitudes 
\cite{Bern} and even for non--supersymmetric QED in a self--dual gauge background
\cite{Dunne}.} 
Note that there are two ways of cutting the 
two--loop
graph and that cancels the original combinatorial factor $1/2$.
 
This explicit analysis expresses the two-loop effective charge as
 \be
\label{emu}
\frac 1{e^2_{\rm two\mbox{-}loop}(\mu) } &=&  \frac 1{e_0^2} + 
\frac 1{4\pi^2} \ln \frac {\Lambda_{\rm UV}}{\mu} - 
\frac 1{4\pi^2} \ln Z_{\rm one\mbox{-}loop}
 \ee
with $ Z_{\rm one\mbox{-}loop} = 1 -  (e_0^2/4\pi^2)  \ln (\Lambda_{\rm UV}/\mu)$. 
Thereby, it relates the 2--loop $\beta$ function to the 1--loop anomalous 
dimension of the matter
field in accordance with the general result \cite{NSVZ}. 
Moreover, in terms of the Wilsonean Lagrangian which for $\mu \gg m$ includes the matter part,  
the whole two-loop result is attributed to the matrix element of the one-loop corrections to 
the matter part \cite{QED}. In this sense, there is no higher-loop corrections to the gauge 
coupling in the  Wilsonean Lagrangian, they  all dwell in the matter $Z$ factor.

The same, of course, follows from the 
analysis of the regime $\mu \ll m$, as we explained in the Introduction: we saw that the 2--loop 
contribution to ${\cal L}_{\rm eff}$
is zero in this case, but the nonzero 2--loop contribution in  $1/e^2_{\rm phys}$ depended
on the physical mass renormalization, which was determined by the renormalization of the kinetic
term related to anomalous dimensions. As a result, we obtained the result (\ref{ephys}) which coincides
with (\ref{emu}) up to the interchange $m_{\rm phys} \leftrightarrow \mu$. 

\subsubsection{Remark on the gauge dependence}
In the derivation above we used the Feynman gauge choice, $\alpha=1$. In a generic
gauge, the propagator  of the vector superfield (\ref{supprop})  involves an extra term 
$\propto (\alpha -1) (D_1^2 \bar D_2^2 +\bar D_1^2  D_2^2) \delta^4 (\theta_1 - \theta_2)/k^4$, 
 which is more singular in infrared than the main $1/k^{2}$ term. 
When $\alpha \neq 1$,  the tadpole diagram, involving the vector field loop and the 
$\bar \Phi \Phi v^{2}$ vertex, does not vanish and  should be added to the two-loop graph (b) 
in Fig.\,\ref{sgraph}.

At the two-loop level it is not difficult to verify that the effective Lagrangian 
does not depend on 
the gauge choice. Extra contributions due to the change of the propagator 
 are canceled out in the sum of two diagrams. 
The situation is a little bit more subtle for  $Z$-factors of the matter fields.
The gauge-dependent term  brings about 
logarithmic infrared divergences there  \cite{Piguet, AGZ}. 
At the one--loop level, they were explicitly evaluated in  \cite{Kovacs}.
 
One may ask now 
 how is it possible to interpret Eq,\,(\ref{emu}) relating the $Z$-factor
to the running charge if this $Z$-factor is gauge--dependent?
However, one can see from explicit calculations  that the gauge-dependent part 
in the $Z$-factor is  due to {\it infrared} range of integration  over virtual momenta, 
 where they are of the same order as external ones. 
 The 
{\it ultraviolet} dependence of the $Z$-factor does not depend on the gauge. 
In other terms, we can formulate this as a statement of 
gauge independence of the {\, Wilsonean} $Z$-factor, 
in which the infrared part should be omitted by construction. 
It is this Wilsonean $Z$-factor which enters the relation 
(\ref{emu}). 

A particular way to calculate
the gauge-independent Wilsonean $Z$-factors is to introduce a nonvanishing mass 
for the quantum vector field $v$ by adding $\int {\rm d}^{4}\theta\, \mu^{2}v^{2}/4e^{2}$ 
to the Lagrangian.  The propagator then becomes
$$
\langle v_1 v_2 \rangle = -2e^2
\int\! \frac {{\rm d}^4 k}{(2\pi)^4}\, {\rm e}^{ik(x_2 - x_1)} \, \frac i{k^2-\mu^{2}} 
\left[1+\frac{\alpha -1}{k^{2}-\alpha \mu^{2}}\, \frac{{D}^2_1 {\bar D}^2_2 +\bar {D}^2_1 {D}^2_2}{16} \right]\delta^4(\theta_1 - \theta_2)
$$
It is sufficient to choose  both $\mu^{2}$ and  $\alpha\mu^{2}$ to be much larger than the virtuality of  external matter line $p^{2}-m^{2}$  to  get rid of the gauge-dependent infrared
part in $Z$.

\subsubsection{Three and higher loops}
\label{sec:3h}

\begin{figure}[h]
   \begin{center}
 \includegraphics[width=4.8in]{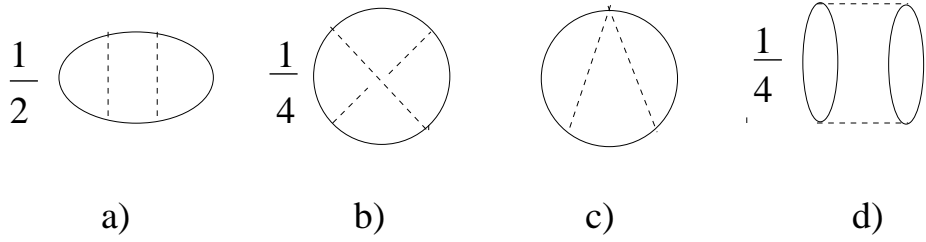}
        \vspace{-2mm}
    \end{center}
   \caption{\small Three--loop effective action.}
  \label{3pet}
  \end{figure}

\begin{figure}[h]
   \begin{center}
 \includegraphics[width=4.8in]{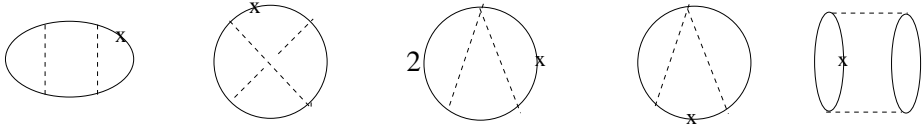}
        \vspace{-2mm}
    \end{center}
   \caption{\small Infrared contributions associated with $\Sigma^{(2)}$.}
  \label{3petsig2}
  \end{figure}

\begin{figure}[h]
   \begin{center}
 \includegraphics[width=2.5in]{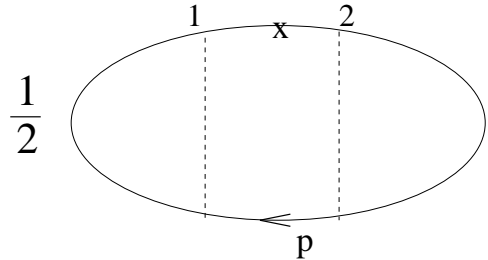}
        \vspace{-2mm}
    \end{center}
   \caption{\small The infrared contribution associated with $\frac 12 [\Sigma^{(1)}]^2$.}
  \label{Sig1kvad}
  \end{figure}

Let us see now what happens at the 3--loop level and higher. 
The relevant 3--loop supergraphs (with the proper combinatorial factors) 
are drawn in Fig.\,\ref{3pet}.
Expanding over $\Gamma_\mu$ produces the d'Alembert operator acting on the matter field
momenta. The infrared contributions shown in Fig.\,\ref{3petsig2} are expressed via the two--loop
contribution to the matter polarization operator $\Sigma^{(2)}$ (it is straightforward to see that
the combinatorial factors come out right). 
Besides, there is an extra infrared contribution coming from the graph in Fig.\,\ref{3pet}a
which is depicted in Fig.\,\ref{Sig1kvad}. Note that the combinatorial factor $1/2$ present
in Fig.\,\ref{3pet}a is not changed here! To understand why, let us look at the 
analytic expression corresponding to this graph. The  singular in momentum $p$ part has the form
  \be
 \label{K2singp}
{\cal K}^{(3)} \ \propto \ \int d^4p \int d^4\theta_2 \, \langle S_1 \bar S_2 \rangle^2 \ .
  \ee
Substituting here the superpropagators from Eq.\,(\ref{supprop}), bearing in mind the relation
(\ref{D2exp}), and expanding the exponential there up to the terms $\propto p^2$,
 we see that the integrand behaves as $1/p^2$ rather than $1/p^4$ at small $p$.
In component language, this means that only the contribution of the {\it fermion} components
of the corresponding superpropagator is relevant such that 
Tr$\left(1 /\, \!\!\!\not\!p \right)^2 \propto 1/p^2$.
Substituting $p_\mu \to p_\mu + \Gamma_\mu$ and expanding over $\Gamma_\mu$, we obtain
the structure $\Box (1/{p^2})$, which is equivalent to inserting a cross in  {\it one}
of the  matter
lines and does not bring about an extra numerical factor.     
Combining all contributions, we obtain
 \be
{\cal K}^{(2)} + {\cal K}^{(3)} + \ldots \ \propto \ \delta Z^{(1)} +  
\frac 12 \left(\delta Z^{(1)} \right)^2 + \delta Z^{(2)}+ \ldots  = 
\ln(1 +  \delta Z^{(1)} +  \delta Z^{(2)} )\ 
 \ee
in accordance with Eq.\,(\ref{ephys1}). 
Similarly, one can separate the infrared contributions in the higher loops.
They give the higher terms of the expansion of the logarithm.

Again, all this was done under assumption that the  gauge is chosen such that
the vector propagator is proportional to $\delta^4(\theta_1 - \theta_2)/k^2$ 
and the coefficient of
  $ (D_1^2 \bar D_2^2 +
\bar D_1^2  D_2^2)  \delta^4(\theta_1 - \theta_2)/k^4$ vanishes (otherwise, the $Z$-factor would
be singular in the infrared). Note that it is not enough now to 
stay in the Feynman gauge $\alpha = 1$. In the latter, the structure  $ \propto (D_1^2 \bar D_2^2 +
\bar D_1^2  D_2^2)  \delta^4(\theta_1 - \theta_2)/k^4$ is absent at the tree level, but is generated
after calculating loop corrections. To cope with this, one should pose
$\alpha = 1 + Ce^2 \ln (\Lambda_{UV}^2/k^2) + O(e^4)$. The dependence of $\alpha$ on $k$ corresponds 
actually to a nonlocal gauge choice \cite{AGZ}. It is not obvious to {\it prove} that everything 
comes out
correctly with this procedure in any order of perturbation theory. We hope to return to this question
in some future work.

\section{Extended {\boldmath ${\cal N}\!=\!2$} SQED: Harmonicity and cancellations}

\subsection{Harmonicity}

As was mentioned in the introduction, the ${\cal N}\!=\!2$ extension of SQED involves an extra 
massless chiral superfield $\Upsilon$,  the following terms are added to the Lagrangian 
(\ref{LSQED}): 
 \be
\label{delLN2}
 \Delta {\cal L} = {\rm Re}\left\{  \frac{1}{2e^{2}}\int {\rm d}^4\theta \, \bar \Upsilon \Upsilon +2
\int {\rm d}^2\theta\,\Upsilon ST  \right\} \ .
   \ee
The mass of the charged matter field is given by a background value of the lowest $\Upsilon$ 
component, $m= \upsilon$.
This explicitly realizes  the moduli nature of the mass. 

Before going over to the ${\cal N} =2$ case, let us discuss an 
interesting feature of 
the one- and two-loop results for the ${\cal N} = 1 $ theory referring to 
their mass dependence.
 At $d\!=\!4$ there is no moduli $\vec A$ and the function 
$h$ in \re{compL} 
depends only on the mass parameter $m$. In the one-loop order $h^{(1)}\propto 
\log |\Lambda_{\rm UV}/m|$ and this dependence can be viewed as a real part of 
$\log (\Lambda_{\rm UV}/m)$ which is an analytic function of $m$. In other words, 
$h^{(1)}$ is a harmonic function in the plane $m_{1}, m_{2}$ which are real and 
imaginary parts of $m=m_{1}+im_{2}$. Both the function and the argument refer to 
the coefficients of the chiral $F$ terms in four dimensions.
 
 However, this holomorphy is broken by higher loops. Indeed,   
$Z$ factor entering Eqs.\,(\ref{ephys1}), (\ref{mrenorm})  is real, and the function
 \be
\label{nonhol}
 \ln \left[ 1 - \frac {e_0^2}{4\pi^2} \left| \ln \frac {\Lambda_{\rm UV}}{m_{\rm phys}}
\right| \right]
 \ee
 is not holomorphic. Only in case of extended ${\cal N}\!=\!2$ 
supersymmetry  $Z$ stays equal to 1,  higher loop contributions in the effective 
charge vanish, and 
 holomorphy is maintained.

Actually, one can {\it derive} that higher loops vanish from the requirement of 
holomorphy, which is a corollary of extended supersymmetry \cite{SeiWit}.
Indeed, in $U(1)$ theory the supersymmetric effective action
 can be expressed in terms of the gauge-invariant ${\cal N} =2$ superfield
  \be
\label{Wscript}
 {\cal W} \ =\ \Upsilon + i\sqrt{2} \tilde{\theta}^\alpha W_\alpha - \frac {\tilde{\theta}^2}4
\bar D^2 \bar \Upsilon
  \ee
as 
 \be
\label{LviaFW}
{\cal L}_{\rm eff} \ =\ \int d^2\theta d^2 \tilde{\theta} \, F({\cal W}) \ + \ {\rm h.c.}  
  \ee
 
The function
$F({\cal W})$ is called prepotential. Doing the integral 
over $d^2\tilde{\theta}$ in Eq.\,(\ref{LviaFW}), we can express the effective
Lagrangian as
  \be
\label{LviaFprim}
{\cal L}_{\rm eff} \ =\ {\rm Re} \left\{ \int d^2\theta  \, F''(\Upsilon)W^2  + 
2\int d^4\theta F'(\Upsilon) \bar \Upsilon\ \right\}\ .  
  \ee
Comparing the first term  
with the corresponding term in \re{LSQED}, we see that 
 $2F''(\Upsilon)$  can be interpreted as the inverse effective
charge $1/e^2_{\rm eff}(\Upsilon)$. By construction,
it is a holomorphic function of the moduli. 

In the region where  $e^2_{\rm eff}(\Upsilon)$ is small, the prepotential can be 
evaluated perturbatively. At the one--loop level, $F''(\Upsilon) \propto \ln \Upsilon$, which is
 holomorphic. The point is that  nonvanishing second or higher loops
 would imply  nonholomorphic dependence
like in Eq.\,(\ref{nonhol}) (with $\Upsilon$ substituted for $m$), which is not allowed. 

In ${\cal N} = 2 $ SQED, the one--loop calculation gives an {\it exact} result for the prepotential.
In non-Abelian theories, there are also nonperturbative  contributions associated with instantons.
These  contributions (they are important in the strong coupling region, 
$\upsilon \sim \Lambda_{\rm IR}$)
 were determined exactly for 
$SU(2)$ theory by Seiberg and Witten \cite{SeiWit}. 
In this paper, we limit our discussion to
perturbative effects.

 Passing to $d<4$ we see from Eq.\,(\ref{h1}) that the harmonicity in $m_{1}, m_{2}$ is lost
 already at the one-loop level.\footnote{\,A similar one-loop phenomenon taking place in four 
 dimensions 
 was discussed in Refs.\cite{KapLou}. Moduli originated from a string construction  led there to
 hierarchical structure of masses.}
 What we have instead, however, is the harmonicity in $2+\hat d$ space where the moduli 
$\vec A$ are added to $m_{1}, m_{2}$, i.e.  harmonicity exists  in the 
extended moduli/parameter space.
 
 Again, looking at the two-loop results we see that this harmonicity is not supported by higher loops.
However, similar to 4d theories, the extended  
 ${\cal N}\!=\!2$ supersymmetry makes harmonicity exact. [See the discussion around Eq.\,(\ref{hN2}). 
We remind that the vector $\vec{A}$ in Eq.\,(\ref{hN2}) involves besides the components 
of vector potential
in reduced dimensions also the lowest component of $\Upsilon$, which
 can be viewed as a linear 
combination $
\upsilon = A_{4}+iA_{5}$.]
This, of course, implies   
vanishing of  higher loops in  ${\cal N}\!=\!2$ theories.

Let us make few more comments on the $d=3$ case, where  the picture is slightly more complicated.
 In three (and obviously also in four) dimensions,
 light degrees of freedom associated with the Abelian 
gauge field come into play. As was explained above, for $d=3$, gauge field is dual to the scalar
one and one obtains an extra moduli, the dual photon $\pi$. 
We have explained before how this duality transformation works 
in the ${\cal N} =1$ case, when restoring
the K\"ahler form of the effective Lagrangian.
In the ${\cal N} = 2$ case, a similar procedure leads to hyper--K\"ahler supersymmetric
$\sigma$ model living on the 
4--dimensional Taub-NUT manifold  with the metric 
  \be
\label{TNUT}
ds^2 = \ h(\vec{A})\, d\vec{A}^2 + h^{-1} (\vec{A}) 
\left( d\pi -  {\small \vecg {\cal A}}(\vec{A})\, d\vec{A} \right)^{2}, \quad \quad
h(\vec{A}) = 1 + \frac {e^2}{4\pi|\vec{A}|}\ ,
 \ee
where 
${\small\vecg {\cal A}}(\vec{A})$
represents the vector potential of an Abelian Dirac monopole
satisfying \\ $\vecg{\partial}\times\vecg{\cal A}=\vecg{\partial} h$.\footnote{\,In the Abelian 
case, 
the result (\ref{TNUT}) is exact. In the non-Abelian $SU(2)$ 
case, a similar expression with $e^2$  substituted by $-2g^2$ describes the asymptotics
of the metric at large $|\vec{A}|$,  whereas the full Atiyah--Hitchin metric involves 
also nontrivial
nonperturbative contributions \cite{SW+CH}.}

The function $h(\vec{A})$   is harmonic, which   
is not accidental. A well--known mathematical
fact is that the K\"ahler potential of a hyper--K\"ahler manifold involving the $U(1)$
isometry is obtained by a Legendre transformation (physically, this is a duality 
transformation)
out of a 3--dimensional harmonic
function \cite{Hitchin}. This harmonic function is nothing but the prepotential
${\cal K}$ in the expression
  \be
{\cal L} \ =\ \frac 1{e^2} \int d^4\theta \, {\cal K} (\vecg{\Gamma}) \ .
  \ee
Now, $\vecg{\Gamma}$ has 3 components, with $\Gamma_3$ representing the superconnection
in the reduced dimension and $(\Gamma_4 + i \Gamma_5)/\sqrt{2} = \Upsilon$. The metric
$h(\vec{A})$ obtained from ${\cal K}$ by $h = -(1/2) \partial^2{\cal K}/(\partial \Gamma_3)^2$
is also harmonic.

\subsection{Cancellations}

The vanishing of two and higher loop contributions to the metric can be confirmed
by direct perturbative calculations.
 In \cite{2loop}, this was done at the two--loop level 
for the $d=1$ theory by evaluating explicitly 
  the relevant graphs: individual contributions to the
effective Lagrangian canceled out in the sum.
 The calculation was done in components and the mechanism for this calculation was not obvious,
however.
We would like to note here that the cancellation becomes transparent if doing the calculations
in the supergraph technique.
 
In the extended case, in addition to the  two-loop graph in Fig.\,{\ref{sgraph}b, there is an 
an extra contribution to the effective Lagrangian due to the  $\Upsilon$ exchange 
depicted by the graph in Fig.\,\ref{Upsilon}.
\begin{figure}[h]
   \begin{center}
 \includegraphics[width=1.0in]{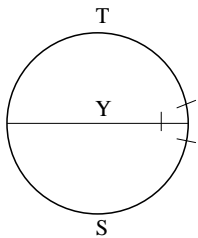}
        \vspace{-2mm}
    \end{center}
\caption{\small Two--loop supergraph with $\Upsilon$ exchange.}
\label{Upsilon}
\end{figure}

\noindent
The 4d $\Upsilon$ propagator has the form
   \be
\label{Upsprop}
  \langle  \Upsilon_{1} \bar \Upsilon_{2} \rangle \ =\  2e^{2}
\int \frac {{\rm d}^dk}{(2\pi)^d}\, e^{ik(x_2-x_1)} \frac i{k^2}\, 
\frac{\bar D_{1}^{2} D_{2}^{2}}{16}\,
\delta^4(\theta_1 - \theta_2) \ .
  \ee
 The differential operators $\bar D_{1}^{2}/4$ and  $D_{2}^{2}/4$ can be absorbed into the vertices 
 completing $\int {\rm d}^{2}\theta_{1}$, $\int {\rm d}^{2}\bar\theta_{2}$ up to $\int {\rm d}^{4}\theta_{1,2}$.
 What is left coincides up to the opposite sign
with the vector field 
 propagator $\langle v_1 v_2 \rangle$ in \re{supprop}. 
 It is straightforward to see then that the contribution of the graph in Fig.\,\ref{Upsilon} 
to ${\cal K}^{(2)}$  exactly cancels the contribution of Fig.\,\ref{sgraph}b.

At the three--loop level, we have now four extra supergraphs depicted in Fig.\,\ref{Upsilon3}.
The full contribution to the effective Lagrangian cancels out, but the mechanism of this
cancellation is not evident as was the case for the two--loop graphs. The situation here is
 similar to what we had at the two--loop level in the component formalism. One should not
be surprised here: the cancellation is the corollary of extended 
${\cal N} = 2$ supersymmetry and 
should not be manifest neither in the component nor in the 
${\cal N} =1$ supergraph formalism. 
We believe that
the cancellation {\it would} become manifest for  any loop if working in the formalism
of  ${\cal N} =2$ harmonic supergraphs \cite{kniga}. 
   \begin{figure}[h]
   \begin{center}
 \includegraphics[width=4.6in]{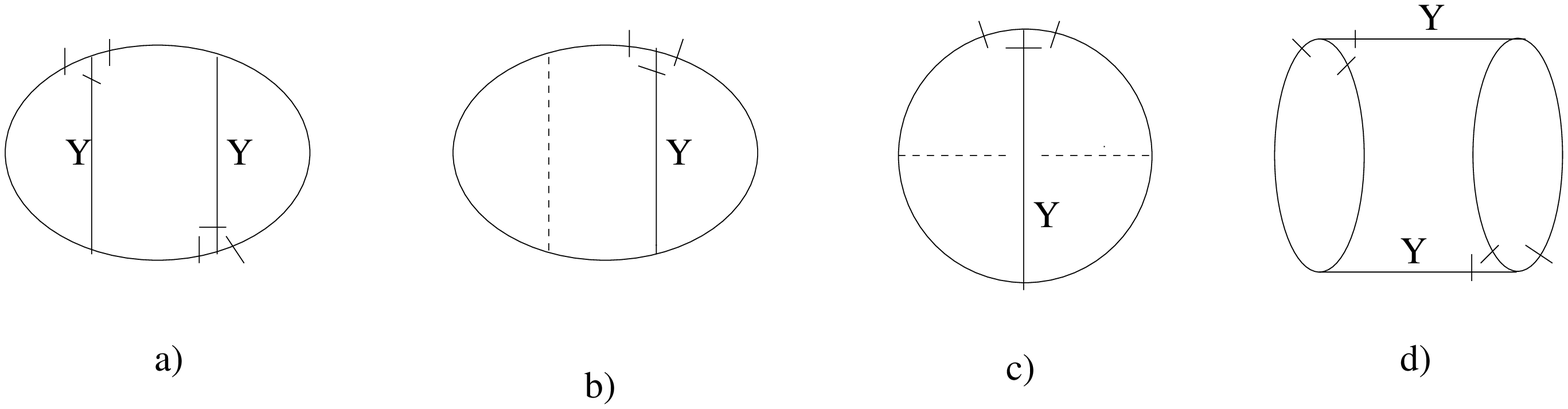}
        \vspace{-2mm}
    \end{center}
\caption{\small Three--loop supergraphs with $\Upsilon$ exchange.}
\label{Upsilon3}
\end{figure}

\section{Non--Abelian theory}

As we discussed in the Introduction, \, ${\cal N} =1$ SYM theory in low dimensions,
$d <4$, involves the moduli associated with Abelian components of gauge potentials in reduced
dimensions. The effective low--energy Lagrangian depends on the corresponding moduli fields
(and also massless gauge fields at $d=3$).
For $SU(2)$ theory they are $A^3_\mu$ and their superpartners $\lambda^3_\alpha$. 
The effective Lagrangian is obtained after integrating out the heavy charged fields
 $A^\pm_\mu,\ \lambda^\pm_\alpha$. The mass of the latter depends on the moduli 
$A^3_{\hat \mu}$. We will see below that the calculations of effective Lagrangian
at the one and two--loop level in supergraph technique are effectively reduced to Abelian
ones. 

 The Lagrangian of the theory has the form
 \be
\label{Lnonab}
{\cal L} \ ={\rm Re} \,\frac{1}{g^{2}}  \int {\rm d}^2\theta\, 
{\rm Tr} \{ W^\alpha W_\alpha \}\ ,
  \ee
where 
$$ W_\alpha = \frac {\bar D^2}8\, e^{-V} D_\alpha e^V \ .$$

 We will restrict ourselves by the case $G = SU(2)$. A earlier, we perform our calculations
in the background field method\cite{Siegel}, i.e. substituting $V \to V+v$, where $V$ is
now a classical background field, which we assume to be Abelian.

We choose the Feynman gauge adding to the  
Lagrangian
the term 
$ - {\rm Tr} \int\! d^4\theta {\nabla}^2 v {\overline \nabla}^2 v/16g^2 $, where 
${\nabla}_\alpha = D_\alpha - i \Gamma_\alpha$ is the covariant spinor derivative.\footnote{For 
non-Abelian theories, the 
study of gauge dependence is more involved  than for SQED, 
and  we will not try here to explore this issue.}
The terms quadratic in $v$ acquire the form
  \be
\label{Lbackv2}
 {\cal L}^{(2)} =\frac 1{2g^2} {\rm Tr} \!\int\! d^4\theta\, v\left[  \nabla_\mu \nabla^\mu
+ iW^\alpha \nabla_\alpha
+ i {\overline W}^{\,\dot{\alpha}}\, {\overline\nabla}_{\dot{\alpha}} \right] v\ .
 \ee
The  covariant d'Alembert operator can be split up as
 $$ 
 \nabla_\mu \nabla^\mu  \ =\  \nabla_{\bar \mu} \nabla^{\bar \mu} + \Gamma_{\hat k}^2
\quad \quad (\bar \mu = 0,\ldots,d-1;\ {\hat k} = d, \ldots,4)\ . 
$$
The moduli $A_{\hat k}$ are the lowest components of $\Gamma^3_{\hat k}$.
They give mass to the charged fields $v^\pm$ and will not be treated perturbatively.

Fixing the gauge leads to appearance of 
 ghosts. The ghosts have the same algebraic nature as the parameters
of gauge transformation, i.e. they are adjoint chiral superfields. Note that
the number of ghost degrees of freedom is two times more than the number of gauge
parameters (for example, for usual Yang--Mills theory, there are $N^2-1$ gauge
parameters and $N^2-1$ {\it complex} ghosts $c$). In supersymmetric case \cite{Siegel,West},
this means that we have two different ghost chiral superfields $c$, $c'$.
On top of this, there
is also the Kallosh--Nielsen ghost $b$, which appears due to the fact that the gauge fixing
term is by itself field--dependent. But the Kallosh--Nielsen ghost
contributes only at the 1--loop level and does not appear in the multiloop graphs.

The calculation of the effective Lagrangian at the one--loop level
is straightforward now. Consider first the loop of the $v$ field. The vertices
of its interaction with the background field $V$  can be read out from
the Lagrangian (\ref{Lbackv2}). They involve at most one spinor derivative. 
Considering the graph with two such vertices  and the 
propagators 
$$\langle v_1^+ v_2^- \rangle \ =\ - \frac {2ig^2}{k^2-\vec{A}^2}\,
\delta^4(\theta_1 - \theta_2)\ , 
$$
 one can observe that the loop vanishes. Indeed,
each propagator involve the factor $\delta^{4}(\theta_1 - \theta_2)$, their product is
zero and two covariant derivatives coming from the vertices are not able to cope with this.

We are left with the ghost loops. To reduce their calculation to that in SQED, 
it is convenient to introduce $G_{f}$, $f=1,2,3$ such that 
$$
G_1 \equiv c\,, \quad G_2 \equiv c'\,, \quad 
\bar G_1 \equiv \bar c'\,,\quad \bar G_2 \equiv \bar c\,, \quad
G_{3}\equiv b\,, \quad \bar G_{3}\equiv \bar b\ .
$$
Only charged ghosts $G_i^\pm, \bar G_i^\pm$, $i =1,2,3$,
 interact
with the external field. Their propagators and interaction vertices are the
same as for the chiral matter fields $S,T$ in massless SQED. Each ghost loop gives the same 
contribution as the SQED matter loops up to a sign.  Thus, the non--Abelian one--loop result is
obtained from the Abelian one [see Eq.(\ref{1loop})], if multiplying it by $-3$ and substituting
$e^2 \to g^2$.
This conforms with the  previous component calculations for the metric 
\cite{Akhmedov,sestry}, 
  \be
  \label{hSU2}
 h_{1d}^{SU(2)}(\vec{A}) &=& 1 - \frac {3g^2}{2|\vec{A}|^3} + \ldots\ , \nonumber \\
  h_{2d}^{SU(2)}(\vec{A}) &=& 1 - \frac {3g^2}{2\pi |\vec{A}|^2 } + \ldots \ ,  \nonumber \\
  h_{3d}^{SU(2)}(A) &=& 1 - \frac {3g^2}{4\pi A} + \ldots \ .  
\ee   

 We are going to show now  
that the same factor $-3$ relates the Abelian and non--Abelian 
contributions to the metric at the two--loop level. 
 The interaction of the ghosts  with the quantum superfield $v^a$ can be derived
by standard  methods. The cubic term
 has the form (see Eqs.\,(6.2.20), (6.2.22) of Ref. \cite{Siegel})
 $$
 {\cal L}_{\rm ghost} \ = \frac {ig}4 \int d^4\theta\  \epsilon^{abc} v^a 
\left(\bar G_1 + G_2 \right)^b \left( G_1 + \bar G_2 \right)^c .
$$

 The relevant two--loop diagrams are drawn in Fig.\,\ref{duhi}. The corresponding analytic
expressions have the same structure (\ref{KGamres}), (\ref{KGamres4}) 
 as in the Abelian case, but the values
of color and combinatorial factors are such that the net contribution of the three graphs
in Fig.\,\ref{duhi} is zero \cite{GZ}.

 \begin{figure}[h]
   \begin{center}
 \includegraphics[width=4.9in]{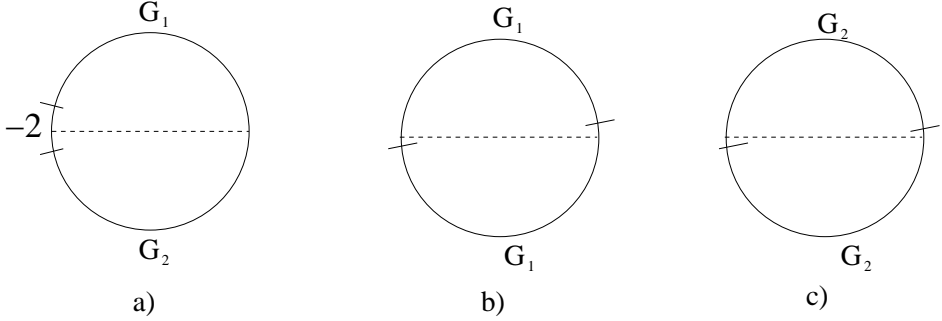}
        \vspace{-2mm}
    \end{center}
\caption{\small Two-loop graphs with ghosts, $-2$ is a relative combinatorial coefficient.}
\label{duhi}
\end{figure}

Thus, we are left only with the graph with three gauge field lines depicted in Fig.\,\ref{3V}.
Expanding (\ref{Lnonab}) in $v$ and ``converting'' one of the factors $\bar D^2$ to 
$-4\int d^2 \bar \theta$, we express the cubic interaction term as
 \be
\label{L3}
{\cal L}^{(3)} = - \frac 1{8g^2} \int d^4\theta\, {\rm Tr} 
\left\{ (\bar D^2 D_\alpha v) (D^\alpha v) v \right \}\ .
 \ee
[Strictly speaking, covariant derivatives $\nabla_\alpha$ enter, but we can substitute them
by $D_\alpha$ by the same token as in the Abelian case; see the footnote after Eq.(\ref{S1S2})].
 
In color vector notations, the
vertex (\ref{L3}) involves the factor $\epsilon^{abc}$ so that
two of the lines are charged with respect to the external field and acquire the mass 
$|\vec{A}|$,
and the third line is neutral and remains 
 massless.
     \begin{figure}[h]
   \begin{center}
 \includegraphics[width=1.5in]{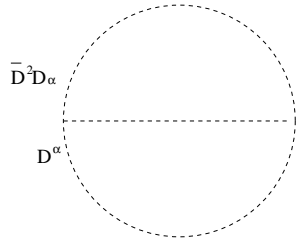}
        \vspace{-2mm}
    \end{center}
\caption{\small Two-loop graph with cubic vector vertices. Covariant derivative 
factors at the left vertex are displayed.}
\label{3V}
\end{figure}

 Note that the vertex has several terms distinguished by the way the factors 
$\bar D^2 D_\alpha$ and $D^\alpha$ are attributed to different lines. Therefore, 
we actually have not one  but several diagrams. 
In principle, one could draw $3! = 6$ such diagrams, but those of them that involve
more that 4 covariant derivative factors on a given line vanish \cite{Siegel,Grisaru}
and we are left with
only four terms written in a symbolic way as
  \be
\label{4terms}
- \!\left(\hspace{-2mm} \begin{array}{c}   \bar D^2 D_\alpha \langle v_1 v_2 \rangle  D^\beta \\[1mm]
D^\alpha \langle v_1 v_2 \rangle   \bar D^2 D_\beta \\[1mm] \langle v_1 v_2 \rangle 
 \end{array}\hspace{-2mm} \right)\! -\!
 \left(\hspace{-2mm} \begin{array}{cc}    \bar D^2 D_\alpha  \langle v_1 v_2 \rangle  \\[1mm]
D^\alpha \langle v_1 v_2 \rangle   D^\beta \\[1mm] \langle v_1 v_2 \rangle 
  \bar D^2 D_\beta \end{array} \hspace{-2mm}\right)\! +\!
  \left(\hspace{-2mm} \begin{array}{cc}   \bar D^2 D_\alpha \langle v_1 v_2 \rangle  D^\beta \\[1mm]
D^\alpha \langle v_1 v_2 \rangle  \\[1mm] \langle v_1 v_2 \rangle 
  \bar D^2 D_\beta  \end{array} \hspace{-2mm}\right)\! +\!
 \left(\hspace{-2mm} \begin{array}{cc}   \bar D^2 D_\alpha \langle v_1 v_2 \rangle  \\[1mm]
D^\alpha \langle v_1 v_2 \rangle   \bar D^2 D_\beta \\[1mm] \langle v_1 v_2 \rangle 
 D^\beta  \end{array} \hspace{-2mm}\right)
  \ee
Each row in the individual term in Eq.\,(\ref{4terms}) corresponds to a propagator
of vector superfield with covariant derivatives acting from the left and from
 the right.

Next, we are using the $D$--algebra rules \cite{Siegel,Grisaru,GZ} , which allow one to 
flip the covariant derivative factors from left to right and back on a given line
and from one line to another at a given vertex. After a simple massaging, 
 all the terms in Eq.\,(\ref{4terms}) are reduced to a standard form
  \be
\label{canon}
 \left( \begin{array}{cc}   \bar D^2 D^2 \langle v_1 v_2 \rangle \\[1mm]
 \bar D^2 D^2  \langle v_1 v_2 \rangle \\[1mm] \langle v_1 v_2 \rangle \end{array} \right)
 \ee
The analytic expression corresponding to Eq.\,(\ref{canon}) is the same as for the 
Abelian graph in Fig.\,\ref{sgraph}b. Collecting accurately all the coefficients, we obtain the net
coefficient $-3$ compared to the Abelian case, as was 
announced.\footnote{\,The result is not new, 
of course. It was obtained in \cite{GZ} by direct
calculation of the  of the effective action in external gauge background. 
Our calculation is much simpler, however. We do not calculate directly the effective
action, but the function ${\cal K}$ such that ${\cal L}_{\rm eff} = \int d^4\theta {\cal K}$.
This allowed us to keep the background very simple - a constant gauge potential. 
The effective action in such a background vanishes, but the function ${\cal K}$ does not.} 

It is curious to observe that in $1d$ theory the metric has the form
   \be
  \label{hnab}
h(\vec{A}) \ =\  1 - \frac {3g^2}{2|\vec{A}|^3} + 
\frac {9g^4}{4|\vec{A}|^6} + \ldots \ ,
   \ee
which coincides with the expansion of
\be
\label{guess}
 \frac 1{1 + \mbox{\large$\frac {3g^2}{2|\vec{A}|^3}$}} \ .
 \ee
We do not find reasons to believe, however, that this reproduces correctly 
also the higher order corrections in the metric.

In ${\cal N} = 2$ non-Abelian theory, 
we should add to the graph in Fig.\,\ref{3V} also the graphs involving
the loop of the adjoint chiral multiplet 
$\Upsilon^a$. Again, the structure of the integral is exactly the same as for the graph
Fig.\,\ref{sgraph}b, but it involves an extra color factor 3. This exactly cancels the 
contribution of the graph in Fig.\,\ref{3V}, as expected.

Going back to ${\cal N} = 1$ theory, note that the coefficient $-3$ is  universal and
appears in any dimension including dimension 4. 
For $d=4$, we do not have moduli and, to determine the one--particle irreducible
effective action in the external background field at scale $\mu$, we should
evaluate the graphs integrating over momenta $ p > \mu$. To determine the 
effective charge (the coefficient of Tr$\{F_{\mu\nu}^2\}$), it suffices to restrict
oneself to an Abelian background. After this, we can repeat
the above reasoning and reduce the task of calculating the non--Abelian graph in 
Fig.\,\ref{3V} to that for the graph in Fig.\,\ref{sgraph}b. The factor $-3$ is thereby 
reproduced. The effective charge is
   \be
\label{gmu}
\frac 1{g^2(\mu)} \ =\  \frac 1{g_0^2} - 
\frac 3{4\pi^2} \ln \frac {\Lambda_{\rm UV}}{\mu} 
- \frac {3g_0^2}{16\pi^4} \ln \frac {\Lambda_{\rm UV}}{\mu} + \ldots \ .
   \ee
This implies 
  \be
 \label{bet2loop}
\beta(\alpha_s) = - \frac {d\, \alpha_s(\mu)}{d\, \ln \mu}\ =\ - \frac {3 \alpha_s^2}\pi 
\left(1 + \frac{\alpha_s}\pi  + \ldots \right)\,,
  \ee
what coincides, of course, with the expansion of the exact $\beta$ function \cite{NSVZ}
    \be
 \label{bettoch}
\beta(\alpha_s) \ =\ -\frac{3 c_V}{2\pi} \frac{\alpha_s^2} 
{1 - \mbox{\large$\frac{c_{\small V}}{2\pi}$} \,\alpha_s }\ .
  \ee

We can give now an interpretation of this result repeating our discussion of Abelian theory
 in the previous
section. An actual calculation of the effective action for $d=4$ in the region $\mu \gg m$
(obviously, non--Abelian theory is massless) requires expansion of the integrand
over $\Gamma_\mu$ and subtracting from $\int$ ({\sl total derivative}) $= 0$ 
the infrared contribution.
This amounts to cutting certain lines and expresses the result via polarization operators
of the corresponding superfields  in lower orders. 
At the two--loop level and working in the Feynman gauge, we observed that the graphs with ghosts
cancel, which implies that the 2--loop $\beta$ function is expressed 
via the 1--loop $Z$-factor of the vector 
superfield, 
  \be
\label{gmuZ}
\frac 1{g^2_{\rm two\mbox{-}loops}(\mu)} \ =\  \frac 1{g_0^2} - 
\frac 3{4\pi^2} \ln \frac {\Lambda_{\rm UV}}{\mu} 
+ \frac 1{4\pi^2} \ln Z^v_{\rm one\mbox{-}loop} \ . 
   \ee
Comparing this with (\ref{gmu}--\ref{bettoch}), we see that $ Z^v_{\rm one\mbox{-}loop}$
just {\it coincides} with the one--loop charge renormalization. Indeed, this was 
the result of the explicit calculation of Refs.\cite{Grisaru}. For sure, it is specific
for SYM in the gauge chosen. In an ordinary gauge theory, renormalization factor of the 
gluon propagator has, generically, nothing to do with the renormalization of the effective
charge. 

Our {\it guess} is that the higher loop diagrams have a similar behavior under condition
that one works
in the nonlocal gauge of Ref.\cite{AGZ} (a refinement of the Feynman gauge -- see the discussion
at the end of Sect.\,\ref{sec:3h}), 
which kills infrared singular contributions in the propagators. Namely, in any order
 {\it (i)}  ghost loops cancel out and  {\it (ii)} renormalization factor of the vector 
propagator coincides with the effective charge renormalization. 
 This expresses the $n$-th contribution
to the $\beta$ function via the $n-1$-th one in the way prescribed by Eq.\,(\ref{bettoch}).    
In terms of operator Wilsonean action, the second and higher loops appear as its matrix
elements. 

\section{Conclusions}
In this paper, we have observed that the corrections  to the effective Lagrangian
of supersymmetric gauge theories
in different dimensions obtained from the 4d theory by dimensional reduction procedure
can be expressed in a closed universal form in the framework of the
supergraph background field formalism, see Eqs.(\ref{1loop},\ref{h2d}). 
For ${\cal N} = 2$ theories, all corrections beyond first loop vanish.
As was discussed in details in Sect.\,3, 
the universal reason for that is the  requirement of {\it harmonicity}
(a generalization
of holomorphy requirement for 4d theories) following from extended supersymmetry.
One--loop corrections are always harmonic, even in ${\cal N} =1$ theories when mass parameters
are considered as extra moduli, but
higher loop corrections are not and so they vanish for ${\cal N} =2$.

Another methodic point where we tried to shed some more light refers to four dimensions and
is the origin of the
exact relation (\ref{ephys1}) expressing higher loop corrections to the effective charge
in supersymmetric SQED via $Z$--factor of charged matter fields. A general proof of it
was discussed in the Introduction and is well known. It is instructive, however, to reproduce
this result by direct calculation of Feynman graphs. We did it in Sect.\,2.5 . It turns out that
the contribution of an arbitrary 
multiloop graph depends on a kinematical region where one of the 
charged field lines goes on shell, and the result depends on a subgraph describing a contribution
to the charged polarization operator.

In Sect.\,4,  we perform a similar calculation for non-Abelian  supersymmetric pure gauge theory.
In that case, the contribution of a diagram describing  an $n$-loop correction 
to the effective charge is expressed via subgraphs describing $Z$--factor of the
gauge field (the only charged physical field in the theory). In other words, 
a recurrent formula exists expressing $n$--loop contribution to the $\beta$ function
via a $n-1$-th one. This leads to the known result (\ref{bettoch}).
We showed it  explicitly at the two--loop level. It would be interesting
to extend this analysis to higher loops.

\section*{Acknowledgments}
We are indebted to Misha Shifman and Misha Voloshin for illuminating discussions 
and to Marc Grisaru for useful correspondence.  A.S.
acknowledges kind hospitality extended to him during his frequent visits to TPI, where
this work was mostly done.       
The work of A.V. was supported in part by DOE grant DE-FG02-94ER408.
  
\pagebreak


{\small
\begin{thebibliography}{99}

 
\bibitem{NSVZ} V.A. Novikov, M.A. Shifman, A.I. Vainshtein and 
V.I. Zakharov,\\ 
Nucl.\ Phys.\ B {\bf 229}, 381 (1983);
Phys.\ Lett.\ B {\bf 166}, 329 (1986).

\bibitem{WB}  J. Wess and J. Bagger, {\it Supersymmetry and 
supergravity}, (Princeton Univ. Press, Princeton,  1983).

\bibitem{Siegel} S.J. Gates, Jr., M.T. Grisaru, M. Ro\^cek and W. Siegel, 
{\it Superspace: or 1001 lessons
in  supersymmetry}, Frontiers in physics: v. 58 (Benjamin/Cummings, 
1983)  [hep-th/0108200].

\bibitem{West} P.C. West, {\it Introduction to supersymmetry and 
supergravity} (World Scientific, 1990).

\bibitem{QED}
M.A.~Shifman, A.I.~Vainshtein and V.I.~Zakharov,
Phys.\ Lett.\ B {\bf 166}, 334 (1986);
M.A.~Shifman and A.I.~Vainshtein,
Nucl.\ Phys.\ B {\bf 277}, 456 (1986).

\bibitem{SV} 
M.A.~Shifman and A.I.~Vainshtein,
Nucl.\ Phys.\ B {\bf 296}, 445 (1988)
[Sov.\ Phys.\ JETP {\bf 66}, 1100 (1987)]; \\
N.~Seiberg,
{\em The power of holomorphy: Exact results in 4-D SUSY field 
theories},\\
hep-th/9408013.


\bibitem{SeiWit} N.~Seiberg and E.~Witten,
Nucl.\ Phys.\ B {\bf 426}, 19 (1994)
[Erratum-ibid.\ B {\bf 430}, 485 (1994)]
[hep-th/9407087].

\bibitem{instsuppot} I.~Affleck, M.~Dine and N.~Seiberg,
Nucl.\ Phys.\ B {\bf 256}, 557 (1985).

\bibitem{sestry} A.V. Smilga, {\it Low--dimensional 
sisters of Seiberg--Witten effective theory},\\  hep-th/0403294,
to be published in the Ian Kogan Memorial Volume. 

\bibitem{LeffN1} A.V. Smilga, 
Nucl.\ Phys.\ B {\bf 291}, 241 (1987);

\bibitem{eshche} E.A. Ivanov and A.V. Smilga, 
Phys.\ Lett.\ B {\bf 257}, 79 (1991);\\ 
A.V. Smilga, 
JHEP {\bf 0204}, 054 (2002)
[hep-th/0201048].

\bibitem{N2} A.V. Smilga, 
Nucl.\ Phys.\ B {\bf 652}, 93 (2003)
[hep-th/0209187].

\bibitem{DE}  D.-E.~Diaconescu and R.~Entin, 
Phys.\ Rev.\ D {\bf 56}, 8045 (1997)
[hep-th/9706059].

\bibitem{DS} D.-E.~Diaconescu and N.~Seiberg,
JHEP {\bf 9707}, 001 (1997)
[hep-th/9707158].

\bibitem{GHR} S.~J.~Gates, C.~M.~Hull and M.~Ro\^cek,
Nucl.\ Phys.\ B {\bf 248}, 157 (1984).

\bibitem{SW+CH} N.~Seiberg and E.~Witten, {\it Gauge dynamics and 
compactification to three dimensions},\\
hep-th/9607163; \\
G.~Chalmers and A.~Hanany,
Nucl.\ Phys.\ B {\bf 489}, 223 (1997)  [hep-th/9608105].

\bibitem{seliv} K.G. Selivanov and A.V. Smilga, 
JHEP {\bf 0312}, 027 (2003)
[hep-th/0301230].

\bibitem{Akhmedov} E.T.~Akhmedov and A.~V.~Smilga, 
Phys.\ Atom.\ Nucl.\  {\bf 66}, 2238 (2003)
[Yad.\ Fiz.\  {\bf 66}, 2290 (2003)]
[hep-th/0202027].


\bibitem{2loop} A.V. Smilga, 
Nucl.\ Phys.\ B {\bf 659}, 424 (2003)
[hep-th/0205044].

\bibitem{Witten:df}
E.~Witten,
Nucl.\ Phys.\ B {\bf 202}, 253 (1982).

\bibitem{scur} M.T. Grisaru, B. Milewski and D. Zanon, 
Nucl.\ Phys.\ B {\bf 266}, 589 (1986).

\bibitem{AGF} L. Alvarez--Gaume and D.Z. Freedman, 
Commun.\ Math.\ Phys.\  {\bf 80}, 443 (1981).

\bibitem{Bern} C. Anastasiou, L.J. Dixon, Z. Bern and D.A. Cosower, 
{\it Cross order relations\\ in ${\cal N} =4$ supersymmetric 
gauge theories}, hep-th/0402053. 

\bibitem{Dunne} G.V. Dunne, 
JHEP {\bf 0402}, 013 (2004)
[hep-th/0311167].

\bibitem{Piguet} O. Piguet and A. Rouet, 
Nucl.\ Phys.\ B {\bf 108}, 265 (1976).

\bibitem{AGZ} L.F. Abbott, M.T. Grisaru and D. Zanon, 
Nucl.\ Phys.\ B {\bf 244}, 454 (1984).

\bibitem{Kovacs} S. Kovacs, {\it A perturbative reanalysis of 
${\cal N} =4$ supersymmetric Yang--Mills  theory},\\ hep-th/9902047.

\bibitem{KapLou}
L.J.~Dixon, V.~Kaplunovsky and J.~Louis,
Nucl.\ Phys.\ B {\bf 355}, 649 (1991);\\
M.~A.~Shifman and A.~I.~Vainshtein,
Nucl.\ Phys.\ B {\bf 359}, 571 (1991).

\bibitem{Hitchin}
N.J.~Hitchin, A.~Karlhede, U.~Lindstrom and M.~Rocek,\\
Commun.\ Math.\ Phys.\  {\bf 108}, 535 (1987).

\bibitem{kniga}
 A.S. Galperin, E.A. Ivanov, V.I. Ogievetsky and E. Sokatchev,\\ 
{\it Harmonic superspace}, (Cambridge, 2001).
 
\bibitem{GZ} M.T. Grisaru and D. Zanon, 
Nucl.\ Phys.\ B {\bf 252}, 578 (1985).

\bibitem{Grisaru}
M.T. Grisaru, M. Ro\^cek and W. Siegel, 
Nucl.\ Phys.\ B {\bf 159}, 429 (1979);
{\bf 183}, 141 (1981).



\end{thebibliography}
}
\end{document}